\documentclass[12pt,preprint]{aastex}
\usepackage{epsfig}
\usepackage{natbib}
\begin{document}
\title{Determining the Lensing Fraction of SDSS Quasars: Methods and Results from the EDR}
\author{
Bart~Pindor\altaffilmark{1},
Edwin~L.~Turner\altaffilmark{1},
Robert~H.~Lupton\altaffilmark{1}, \&
J.~Brinkmann\altaffilmark{2}
}
\altaffiltext{1}{Princeton University Observatory, Princeton, NJ 08544, pindor,elt,rhl@astro.princeton.edu}
\altaffiltext{2}{Apache Point Observatory, 2001 Apache Point Road, P.O. Box 59, Sunspot, NM 88349-0059,jb@apo.nmsu.edu}  

\bibliographystyle{aj}

\begin{abstract}

We present an algorithm for selecting gravitational lens candidates from amongst Sloan Digital Sky Survey (SDSS) quasars. In median Early Data Release (EDR) conditions, the algorithm allows for the recovery of pairs of equal flux point sources down to separations of $\sim 0{\farcs}7$ or with flux ratios up to $\sim$ 10:1 at a separation of $1\farcs5$. The algorithm also recovers a wide variety of plausible quad geometries. We also present a method for determining the selection function of this algorithm through the use of simulated SDSS images and introduce a method for calibrating our simulated images through truth-testing with real SDSS data. Finally, we apply our algorithm and selection function to SDSS quasars from the EDR to get an upper bound on the lensing fraction. We find 13 candidates among 5120 z $>$ 0.6 SDSS quasars, implying an observed lensing fraction of not more than 4 $\times 10^{-3}$. There is one likely lens system in our final sample, implying an observed lensing fraction of not less than $3 \times 10^{-5}$ (95$\%$ confidence levels). 

\end{abstract}
\keywords{ quasars: general; gravitational lensing}

\section{Introduction}

The discovery of new gravitationally-lensed quasars, where the gravitational potential of a foreground galaxy produces multiple images of a background quasar, is of broad astrophysical interest. Such systems can be used individually to directly model the mass distribution of the lens galaxy (Kochanek 1995; Cohn et al. 2001; Mu{\~ n}oz, Kochanek, \& Keeton 2001; Koopmans \& Treu 2002; Rusin et al. 2002; Treu \& Koopmans 2002; Trott \& Webster 2002 \nocite{1995ApJ...445..559K} \nocite{2001ApJ...554.1216C} \nocite{2001ApJ...558..657M} \nocite{2002ApJ...568L...5K} \nocite{2002MNRAS.330..205R} \nocite{2002ApJ...575...87T} \nocite{2002MNRAS.334..621T}), investigate dark matter subtructure \citep{2001ApJ...563....9M,2002ApJ...565...17C,2002ApJ...572...25D}, measure the Hubble constant through time delay measurements \citep{1964MNRAS.128..307R,1997ApJ...482...75K,1999ApJ...527..513K} and to probe the source size through microlensing studies \citep{2000MNRAS.315...62W}. Additionally, the statistics of an observed sample of gravitational lenses can be used to infer the mass profiles of the population of lens galaxies \citep{2001ApJ...561...46K,2001ApJ...549L..33R,2001ApJ...563..489T} or to constrain the value of cosmological parameters, in particular the value of the cosmological constant (\nocite{1990ApJ...365L..43T, 1990MNRAS.246P..24F, 1996ApJ...466..638K, 2002ApJ...575L...1K,2002PhRvL..89o1301C,1999A&A...350....1H}Turner 1990; Fukugita, Futamase, \& Kasai 1990; Kochanek 1996; Helbig 1999; Chae et al. 2002; although see Keeton 2002). This last application is of particular relevance in light of the recent evidence of a high value of $\Lambda$ from measurement of high-redshift supernovae \citep{1999ApJ...517..565P,2001ApJ...560...49R}, CMB anisotropy measurements \citep{2002ApJ...571..604N,2002ApJ...568...38H,2001ApJ...561L...1L} and so-called concordance models \citep{1999Sci...284.1481B}. Lensing statistics will almost certainly never be competitive with the accuracy of the aforementioned experiments in terms of measuring the values of cosmological parameters. Nevertheless, confirmation of the high lensing rates predicted in high $\Lambda$ model cosmologies is an important observational test, particularly as the systematic errors for lensing are distinct from those of supernovae experiments.

	The Sloan Digital Sky Survey \citep[henceforth SDSS]{2000AJ....120.1579Y} dataset offers an opportunity to discover a large number of new gravitationally-lensed quasars. Primarily, it is the size of the survey which makes it a superlative dataset for finding lenses. Assuming a plausible lensing fraction of $\sim 10^{-3}$, the complete spectroscopic sample of $10^5$ SDSS quasars will include of order 100 lens systems. Further, the $10^4$ square degrees of imaging data should contain of order $10^6$ quasars within one magnitude of the spectroscopic limit, although their lensing rate will be probably be considerably lower, depending on the details of the quasar luminosity function. A number of other features of the survey are also pertinent. Since the imaging data are taken in five bands nearly simultaneously, it is possible to use colour information and require consistency between different bands to considerably improve the efficiency of most selection techniques for lenses. Further, the SDSS is particularly well-suited to the collection of a statistically well-defined sample of lenses because the uniformity of the data and of the reductions should allow for an accurate determination of the selection function. In this paper we present the methods by which we intend to assemble such a sample as well as the techniques we will use to measure our selection function through the use of simulated images. At present, we consider only the problem of recovering lenses from a sample of objects which are already known to contain at least one quasar component from their SDSS spectra. We will not explicitly treat the problem of how the SDSS quasar target selection may influence the selection function \citep[henceforth K91]{1991ApJ...379..517K}, although we do discuss this problem in the following section.     

When discussing lensing statistics, we define lens fraction to mean the true, unknown, fraction of quasars in the sample which are multiply-imaged and we define observed lens fraction to mean the fraction of quasars in the sample which have been observed to be multiply-imaged. The observed lensing fraction is related to the lensing fraction through the selection function. Note that, even with complete knowledge of the selection function, the lensing fraction cannot be recovered from the observed lensing fraction as there may always be a glut of unobserved lenses in the selected-against regions of parameter space. However, given a theoretical model of the lensing population and cosmology, one can use the selection function to predict the observed fraction and compare to the experimental value.   

Readers unfamiliar with the SDSS are encouraged to consult the SDSS EDR paper \citep{2002AJ....123..485S} in case of encountering unfamiliar SDSS terminology.

The structure of this paper is as follows. In section 2 we briefly consider the problem of selecting lensed quasars in general and discuss those aspects particular to the SDSS. We then introduce a selection algorithm based upon fitting the observed point spread function (henceforth PSF) to the SDSS atlas images, and illustrate how the selection function of this algorithm can be measured. In section 3 we apply our selection technique to a sample of SDSS quasars to show how a sample of lens candidates for high-resolution follow-up can be defined. In section 4 we discuss our results and consider future work. 

\section{Selection Techniques and Selection Functions}

\subsection{General Considerations}

A sample of SDSS lenses will be of greatest interest if it can be selected by a method which could be applied uniformly to the entire SDSS dataset. In an ideal lens survey, every SDSS quasar would be followed-up with high-resolution imaging to test for the presence of multiple images. However, given that the complete spectroscopic sample of SDSS quasars is scheduled to comprise some $10^5$ objects, such an approach is not immediately feasible. An obvious alternative is to use available morphological information from the SDSS imaging data when selecting candidates for follow-up \citep{1988AJ.....95...19W}. In this way the completeness of the survey is reduced, but the efficiency is expected to be greatly increased. Before describing the mechanics of our morphological criterion, we briefly review some of the theoretical expectations and observational constraints which motivate our approach.
   
The precise details of the predicted distribution of lens image splittings depend upon the mass model and luminosity function chosen for the population of lens galaxies. However, it is a generic prediction of realistic models that the peak of the distribution is at about 1$^{\prime\prime}$ and that the majority of systems will exhibit splittings less than 2$^{\prime\prime}$ (Turner, Ostriker, \& Gott 1984, Hinshaw \& Krauss 1987\nocite{1984ApJ...284....1T}\nocite{1987ApJ...320..468H}). Results from the CLASS radio lens survey are in good general agreement with such expectations \citep{Browne02}. Yet the typical FWHM for SDSS Early Date Release \citep{2002AJ....123..485S} imaging data is not better than 1$\farcs 6$ and since the SDSS pixel scale is 0$\farcs4$, critical sampling requires that seeing not be significantly sub-arcsecond. Consequently, to recover a large fraction of lens systems, it is necessary to be able to identify unresolved systems. Although the SDSS photometric pipeline \citep[henceforth PHOTO]{2001adass..10..269L} does include software which deblends compound objects, we find that few pairs with splittings below 3$^{\prime\prime}$ are deblended automatically by the pipeline. In this work we concentrate on methods for recovering such small-splitting lenses. 

For wider separation lenses ($\Delta \theta > 3^{\prime\prime}$), we expect components to begin appearing as separate objects in the SDSS catalogs, and a search for such lenses can be performed by matching pairs in the quasar target lists. It should be noted that fibre collisions prevent the SDSS from obtaining separate spectra on a single plate for pairs of objects with a separation of less than 55$^{\prime\prime}$. There is a considerable amount of overlapping area between different plates ($\sim$40$\%$), so that two nearby objects could be observed on separate plates, but many such close targeted quasar pairs will have only one SDSS spectrum, meaning that even at these separations most lenses cannot be identified from SDSS data alone.   

In addition to separation, the multiplicity and configuration of lens images also greatly affects their detectibility. Almost half (10 of 22) of the lenses found by the CLASS survey exhibit four or more images. Ideally, our selection techniques would be effective at finding lenses of any configuration. However, there are two main reasons why this is presently not the case; i) The computational load associated with techniques which might be best suited for finding higher multiplicity systems (ie additional model components, deconvolutions) begins to become prohibitive for the thousands of objects we wish to analyze. Our morphological criterion (described below) is undoubtedly more sensitive to two than four image systems, ii) The modeling of quads is obviously more involved and less certain than the modeling of two-image systems. In fact, \citet{2001ApJ...553..709R} argue that simple elliptical models cannot reproduce the quad fraction observed by CLASS. As a result, the selection function for quads is difficult to characterize. In section 2, we explicitly calculate the selection function for pairs of point-sources but we are only able to roughly estimate the detectibility of quads. 

In the observational section of this paper, we will only consider selecting lenses from a parent sample of objects which have already been confirmed to contain at least one quasar component by SDSS spectroscopy. This is largely because in our early efforts to select lenses from the photometeric data alone we found a high number of false positives, in particular pairs of blue compact emission-line galaxies which are quite difficult to distinguish from quasars on the basis of broad-band colours alone. However, having effectively restricted ourselves to the SDSS quasar sample, an important question, as pointed out in K91, is whether the selection of the parent quasar sample is biased with respect to lenses. Even if we only consider the quasar images themselves, we already find that the answer is yes. The SDSS quasar target selection \citep[henceforth R02]{2002AJ....123.2945R} effectively requires \footnote{Extended quasars at redshifts greater than 3 may appear in the quasar sample if they are targeted as FIRST sources, ROSAT sources, or if there is a large error in the photometric redshift estimate.} that quasars with a redshift greater than 3.0 be classified as point sources by PHOTO. While this does not eliminate all recoverable lenses, as will be shown in the next section, it does represent a substantial bias. Additional biases can be introduced by the lensing galaxy. The lensing galaxy can cause a lens system to be excluded from the quasar sample either because i) the observed colours of the lens system are changed by flux from the lensing galaxy, ii) the observed shape of the lens system is changed by flux from the lensing galaxy, or iii) the observed colours of the lens system are changed by reddening of the quasar images along different sight-lines in the lensing galaxy.
  
For most lenses, we expect the lensing galaxy to lie at moderate redshift. \citet{1984ApJ...284....1T} find that the peak of the redshift distribution of lenses lies at $z \sim 0.5$, with the majority of lenses lying at redshifts greater than 0.5. For high $\Lambda$ cosmologies, the lenses are expected to lie at moderately higher redshifts (cf. Fukugita et al. 1990\nocite{1990MNRAS.246P..24F}) This result is fairly model-independent since the term involving the angular-diameter distances in the expression for the critical surface density

\begin{eqnarray}
	 \Sigma_{cr} \propto \frac{D_s}{D_d D_{ds}} 
\end{eqnarray}

(e.g. Scheider, Ehlers, \& Falco 1992\nocite{1992grle.book.....S}) always tends to make lensing most likely when the lens lies roughly mid-way between the source and the observer. We can get a feel for the contribution of such a lens if we consider that the expected SDSS $i$ magnitude of an $L_*$ galaxy at $z$ = 0.5 is 19.6 (20.6 and 22.3 in $r$ and $g$, respectively), using $M_{g}* = -20.04$ \citep{2001AJ....121.2358B}, $K$-corrections and colours from \citet{2001AJ....122.2267E}, and including luminosity evolution in the amount $\frac{d\mathrm{log}(M/L)}{dz} = -0.6$ (e.g. \nocite{2001ApJ...553L..39V} van Dokkum et al. 2001) with $\Omega_M = 0.3, \Omega_{\Lambda} = 0.7$ and $H_0$ = 75 km s$^{-1}$ Mpc$^{-1}$. For low redshift quasars ($z<3$) SDSS quasar target selection requires that the $i$ magnitude of an object be less than 19.1 in order to be included in the spectroscopic sample (R02), hence we expect the lensing galaxy to contribute significant flux to the SDSS atlas image in $i$, but less so in the bluer bands. In particular, the $u-g$ colour, the chief identifier of low-redshift quasars, of the low-redshift sample will not be significantly affected by lens galaxy flux. However, for the fainter, higher-redshift sample ($z>3$) the $i$ magnitude limit is 20.2, and the objects are chiefly detected in $i$ and $z$, so the flux from the lens galaxy will likely make a substantial contribution. On the other hand, lensed quasars at higher redshifts will tend to have higher redshift, hence less prominent, lensing galaxies. It should also be noted that the above galaxy magnitudes are in Petrosian magnitudes, while quasar target selection uses PSF magnitudes. This means that target selection will preferentially consider concentrated quasar light over extended galaxy light when making colour cuts, reducing contamination from the lens galaxy. Also, quasar target selection does not simply select objects with 'normal' quasar colours, but rather targets objects which are outliers from the stellar locus. Hence, in order to cause a lens system to be de-selected, flux from the lens galaxy must not only change the colours, but must also change them in such a way that the object will move onto the stellar locus (or into one of the other excluded areas, see R02). The extent to which this is possible depends largely upon the redshifts of the quasar and the lens galaxy (cf. Figure 4 of R02 to see the location of the galaxy locus in SDSS colour space).   

A similar situation exists with regard to the effect of the lens galaxy upon the observed morphology. For the brighter, low-redshift sample, lens galaxy flux will not remove an object from the sample as no morphological cut is made in quasar target selection. For higher redshift objects, only unresolved objects will enter the sample, so systems where the lensing galaxy is prominent will again be excluded. 

Corrections for both of these selection effects must be made before any sample of lenses chosen from the quasar sample can be used for statistical purposes. We would argue that the use of simulated images as described in the next section is the best way to determine the required corrections. However, in this work we will not address this problem explicitly and henceforth we make the simplifying assumption that the flux from the lens galaxy is negligible. Of course, there is a particularily interesting class of lenses for which this assumption will fail quite badly, perhaps most notably for 2237+0305 \citep{1985AJ.....90..691H}. However, the problem of finding lens systems for which the lens galaxy lies at low redshift is one which is possibly best approached through the use of the SDSS main galaxy sample \citep{2000MNRAS.319..879M}.
 
	\citet[henceforth F99]{1999ApJ...523..617F} studied the effects of reddening in known lens systems and estimated the mean extinction to be $E(B-V)$ = 0.10 $\pm$ 0.08 mag. This amount of reddening is comparable to the width of the low-redshift quasar locus in color space (cf. figure 5 of Richards et al. 2001\nocite{2001AJ....121.2308R}), and so should not remove a large fraction of lenses from the quasar sample. In section 3, we present a selection technique which requires consistent colours between lens candidate components while allowing for differential reddening. 

	In summary, none of the effects associated with the lens galaxy are predicted to remove a large fraction of lenses at redshifts less than 3, where quasars will be significantly brighter than the lens galaxies, at least in the bluer bands, and where quasar target selection is unbiased with respect to object morphology. For the higher-redshift sample, flux from the lens galaxy will produce a considerable anti-lens bias both due to colour contamination and altered morphology.    
	
\subsection{Morphological Selection}

	In this and following sections we present results based largely upon simulated SDSS images. The details of how these simulations are produced are found in the appendix. 

	We devised a selection technique for selecting non-point source morphologies based upon fitting the observed PSF to the SDSS atlas images. Specifically, we fit each object twice; once with a model consisting a single PSF, and then again using a model consisting of two independent PSFs. Each PSF is allowed to vary in position and flux (normalization), so that the first model has three parameters and the second has six. The fits are performed using the empirical PSF as measured by the SDSS photometric pipeline. More accurately, since the photometric pipeline measures PSF variation across the frame, the PSF used is one re-constructed according to the object's position (see Lupton et al. 2001\nocite{2001adass..10..269L} for details). For each of these models, we minimize the reduced chi-squared:

\begin{eqnarray}
\chi^2 = \frac{1}{N} \sum_i^N \frac{(x_i - y_i)^2}{(x_i + sky) / gain + dark}
\end{eqnarray}

where $x_i$ are the object counts, $y_i$ are the model counts, $sky$ is the measured sky value, $dark$ is measured dark variance, and $N$ is a sum over pixels in the atlas image. Note that the counts in SDSS atlas images are less the measured sky value. The minimizations are performed using the downhill simplex method \citep{1992nrca.book.....P}. We seed the two-component fit by measuring the inverse-radius weighted moments ($1/r^2$) (cf. R. H. Lupton, in preparation) of the atlas image and taking as our initial guess an equal flux pair with the same moments. We can also measure the signal-to-noise (S/N) for an object as 

\begin{eqnarray}
\mathrm{S/N} = \sum_i^N \frac{x_i}{\sqrt{(x_i + sky) / gain + dark}} 
\end{eqnarray}

By summing over the entire atlas image we do not get optimal signal-to-noise (in the sense of, for instance, Naylor 1998\nocite{1998MNRAS.296..339N}), however, our measure has the advantage of being independent of whether we are fitting a one or two component model. For simulated data the fits presented are in a single band, whereas for the real SDSS atlas images we performed fits independently in $g,r$ and $i$, and, unless otherwise stated, the results of the fits are presented as averages over these three bands weighted by the square of the S/N.

	Our selection algorithm is based upon the difference in the residuals of our one and two component fits. Specifically, if we define $\chi^2_{1}$ and $\chi^2_{2}$ to bethe best-fit value of $\chi^2$ for the 1 or 2 component model, respectively, then our morphological criterion depends upon the statistic $\Delta \chi^2$ where

\begin{eqnarray}
	\Delta \chi^2 \equiv \chi^2_1 - \chi^2_2 
\end{eqnarray}

Clearly, for a noise-free point source, we expect $\Delta \chi^2 = 0$. We chose a threshold value of $\Delta \chi^2$ which we determined would avoid any false positives from actual point source due to Poisson and sky noise. This value was obtained by fitting to, and hence, measuring the fluxes of, 5120 z$ > $0.6 SDSS quasars and then simulating a sample of 5120 equivalent point sources with the same fluxes, seeing, and observed sky values. Figure 1 shows the distribution of $\Delta \chi^2$ for this simulated sample. A Gaussian fit to this distribution predicts that even for the complete sample of $10^5$ quasars, the expected number of false positives is less than one for $\Delta \chi^2 \geq 0.05$, although it is clear that such a fit does somewhat underestimate the number of objects at larger values of $\Delta \chi^2$. We conservatively set our threshold value to be $\Delta \chi^2 \geq 0.1$, and expect no false positives due to photon and sky noise. It should also be noted that of our simulated 5120 objects, 4 did show $\Delta \chi^2 \geq 0.1$, but these events were associated with numerical issues in the fit and converged to lower $\Delta \chi^2$ if the simplex fits were restarted. 

\subsection{Characterizing the Selection Function}

\subsubsection{Two-Component Systems}

In order to characterize the selection function which results from the selection criterion described above, we simulated an ensemble of SDSS images with two point-source objects of varying separation and flux ratio. We sampled the separation direction every 0.2 arcseconds between separations of 0.2 and 3.2 arcseconds and we sampled the flux ratio direction every factor of 2 between 1 and 64. At every lattice point in the separation - flux ratio parameter space we produced ten realizations of the geometry with random orientations, displacements relative to the pixel grid, and realizations of the noise. We also folded-in variations in the observed sky in the ten iterations (see appendix). For our fiducial set of simulations, we assigned a total flux equal to the median total flux in $r$ observed in our sample of 5120 SDSS quasars (corresponding to a psf magnitude of $r \sim$ 19.0) and we set the seeing to be equal to the observed median seeing (1$\farcs$6 arcseconds FWHM, also in $r$) for this same sample. Note that when changing the flux ratio of a pair of point sources we conserved the total flux between them such that all simulated objects would appear identical if completely unresolved. 

Figure 2 shows the regions of the separation - flux ratio parameter space in which objects pass our selection criterion. Note that failure to select a object can either result from the $\Delta \chi^2$ in fact being less than 0.1 for the correct geometry, or from the two-component fit failing to find the correct geometry and corresponding global maximum of $\Delta \chi^2$. This is the case at larger separations and flux ratios (eg 3$\farcs$0 and 4:1). Such failures are due to our seeding of the fit which currently tends to favour small-separation geometries. 

Figure 3 shows the region in which the PHOTO star-galaxy separator identified objects as galaxies. This figure is of particular interest since, as mentioned above, SDSS quasar target selection rejects high-redshift quasar candidates which are classified as galaxies. The actual median flux for high-redshift quasars will be somewhat less than that of the entire sample, meaning that fewer resolved objects would be classified as galaxies, but clearly the bias against lenses is still quite strong. Figure 4 shows the region in which the PHOTO deblends the objects into two separate components. This figure shows the accessible regions of the separation - flux ratio parameter space if one wishes to recover wide-separation lenses from the SDSS catalogs alone. In these median conditions, even at separations of 3$\farcs$2 the PHOTO deblender is only just beginning to deblend point sources with flux ratios less than one. Although we concentrated on smaller separation systems, Figure 5 shows, for reference, the performance of the PHOTO deblender at larger separations. It is apparent that 4$^{\prime\prime}$ is the approximate splitting at which the PHOTO deblender is complete down to flux ratios which can also be recovered by the $\Delta \chi^2$ method. 

	Apart from the value of the $\Delta \chi^2$ statistic, the best-fit two component model also returns a predicted geometry for the system. We can use the known geometry of our simulated images to determine the accuracy of the recovered geometry. Figure 6 shows the best-fit separation as a function of input separation for all objects with $\Delta \chi^2 > $ 0.1 (regardless of flux ratio). Apart from a modest systematic tendency to over-estimate the separation of small separation pairs, the separation is fit quite accurately. Simply taking the standard deviation of the residuals about each point, we estimate the typical error in the separation of be $\simeq$ 0$\farcs$06. Figure 7 shows the logarithmic error in the recovered flux ratio, again for all objects with $\Delta \chi^2 > $ 0.1 . For each input flux ratio we fit a Gaussian to the observed distribution of errors. Although in each flux ratio bin there are a few obvious outliers, these Gaussian fits are good enough that we can estimate the errors directly from the measured $\sigma$. Note that the errors in the 1:1 flux ratio bin are asymmetric because the flux ratio is less then one by construction.

We should be careful to note that since the input flux is randomly assigned from a Poisson distribution, the purported input flux ratios are only accurate to the level of the Poisson noise. Specifically, if $\mathcal{F}_1$ and $\mathcal{F}_2$ represent the fractions of the total flux in the two components, then the uncertainty in the flux ratio from Poisson noise is

\begin{eqnarray}
	\Delta \left(\frac{\mathcal{F}_1}{\mathcal{F}_2} \right) = \left(\frac{\mathcal{F}_1}{\mathcal{F}_2} \right) \frac{1}{\sqrt{\mathrm{(Total Counts)} \mathcal{F}_1 \mathcal{F}_2}} 
\end{eqnarray}

or 

\begin{eqnarray}
	\Delta \mathrm{log} \left(\frac{\mathcal{F}_1}{\mathcal{F}_2} \right) = \frac{1}{\sqrt{\mathrm{(Total Counts)} \mathcal{F}_1 \mathcal{F}_2}} 
\end{eqnarray}

For the median flux, the total counts are $\sim$ 18000, and, for the flux ratios we are considering, $\sqrt{\mathcal{F}_1 \mathcal{F}_2} \sim 0.5$, so that

\begin{eqnarray}
	\Delta \mathrm{log} \left(\frac{\mathcal{F}_1}{\mathcal{F}_2} \right) \sim 0.01 
\end{eqnarray}

Hence the contribution, in quadrature, of the Poisson noise is small enough that we can retain our estimate of the error in the logarithm as simply being the values of $\sigma$ shown in figure 7. 

The accuracy of the recovered flux ratios is of vital importance if we wish to use measured colour differences between lens components as a selection criterion. We define, for example, $D(g-r)$ as the measured $g-r$ colour difference between two lens components. Note that the colour difference, $D$, differs physically from the differential reddening, $\Delta E$, in that it may be caused by intrinsic colour differences, as in the case of a QSO plus star pair, as well by reddening. In our notation $\Delta D$ refers to the error in $D$.    
Then, if $f_g$ and $f_r$ are the measured flux ratios of two lens components in those bands, we have

\begin{eqnarray} 
	D(g-r) = -2.5 \mathrm{log}\left(\frac{f_g}{f_r}\right) 
\end{eqnarray}

so that the error goes as 

\begin{eqnarray} 
	\Delta D(g-r) = 2.5 \frac{ \Delta \left(\frac{f_g}{f_r}\right)}{\frac{f_g}{f_r}}  
	 = 2.5 \sqrt{\left(\frac{\Delta f_g}{f_g} \right)^2 + \left(\frac{\Delta f_r}{f_r} \right)^2} 
\end{eqnarray}

If we take as a typical value of $\frac{\Delta f}{f} = \sigma \simeq $ 0.03, then we get a typical error in the colour difference of

\begin{eqnarray} 
	\Delta D(g-r) \simeq 0.1 
\end{eqnarray}
 
In selecting lens candidates from real data, we will also want consider the consistency of the point source component positions for the best-fit models across the three bands. Even if the flux ratios change between bands, we would still expect the positions of a pair of point sources to be the same if they are being successfully recovered in each of the bands. Further, it is self-evident that consistent positions are a pre-requisite for the meaningful measurement of colours. We define a measure of the difference in component positions between two bands as the minimum total distance that the positions of point source components as fit in first band must be displaced so as to coincide with the positions of the components as fit in the second band, so that

\begin{eqnarray} 
	\Delta \mathrm{position}_{1,2} \equiv \sqrt{(x_{1A} - x_{2A})^2 + (x_{1B} - x_{2B})^2 + (y_{1A} - y_{2A})^2 + (y_{1B} - y_{2B})^2} 
\end{eqnarray}

where the components A and B are chosen so as to minimize the difference. This measure considers changes in both separation and position angle, but is independent of changes in the flux ratio. In order to measure this difference for our simulated objects, we produced two more sets of simulated pairs of point sources which have identical positions to the set shown in figure 2 but which have different realizations of the Poisson noise. We thus have three realizations of each object and we measure the average of the position differences between them

\begin{eqnarray} 
	\Delta \mathrm{position} \equiv \sqrt{(\mathrm{position}_{1,2}^2 + \mathrm{position}_{1,3}^2 + \mathrm{position}_{2,3}^2)/3} 
\end{eqnarray}

Figure 8 shows the distribution of $\Delta \mathrm{position}$ for those objects in which all three realizations have $\Delta \chi^2 > $ 0.1 . When selecting objects on consistent geometry, we will require $\Delta \mathrm{position} < $ 0.4 arcseconds. 
	
	Apart from geometry, the total flux of a lens and the seeing conditions under which it was observed also greatly affect its detectibility. Figure 9 reproduces the fiducial set of simulations shown in figure 2 and illustrates the effect of better and worse seeing as well more or less total flux. The seeing values displayed correspond to the 25th-percentile, median, and 75th-percentile observed seeing (1$\farcs$8,1$\farcs$6,1$\farcs$4), while the total flux values correspond to one magnitude more than the median, the median, and one magnitude less than the median. It should be noted that the median observed seeing in the EDR will probably end up being  good deal worse than the median value for the whole survey. 
	Ultimately, for statistical purposes, we envision a suite of simulations which sufficiently-finely samples the four-dimensional separation - flux ratio - total flux - seeing parameter space to allow the selection function for every quasar in the sample to be accurately determined through simple interpolation. 

\subsubsection{Four-Component Systems}

	While two-component geometries are spanned in the separation - flux ratio parameter space, the equivalent parameter space for four-component geometries has eight dimensions, albeit with some rotation and reflection symmetries. Consequently, not only is an exhaustive simulation of four-component geometries computationally-daunting, it would also be challenging to interpret the results of such a simulation (eight-dimensional contour plots!). Further, as mentioned above, Rusin and Tegmark argue that simple lens models cannot reproduce the observed quad fraction for CLASS lenses, meaning that producing a believable ensemble of lens geometries based on theoretical expectations is at present a dubious undertaking. Nonetheless, we wish to at least qualitatively ascertain how well our morphological selection works for four-component systems. These considerations lead us to the following observationally-motivated toy model. We envision a one-dimensional sequence of quad-geometries running for highly-symmetric to highly-asymmetric. We chose the known lenses H1413$+$117 \citep{1988Natur.334..325M} (symmetric) and PG1115$+$080 \citep{1980Natur.285..641W} (asymmetric) as the end-points of our sequence. If we denote a quad as the sum of four point-sources in the form 

\begin{eqnarray} 
	\mathrm{Q} = \sum_i^4 \left( \begin{array}{c}
				x_i \\
				y_i \\
	 			F_i
				\end{array} \right) 
\end{eqnarray}

where $x$ and $y$ represent the positions, and F represents the flux ratio of  component relative to the primary, then we can produce a sequence of quads by interpolating between our chosen endpoints so that an interpolated quad will have a geometry given by 

\begin{eqnarray} 
\mathrm{Q} = \sum_i^4 a \left( \begin{array}{c}
				x_i (\mathrm{1413})\\
				y_i (\mathrm{1413})\\
	 			F_i (\mathrm{1413})
				\end{array} \right) + \sum_i^4 (1-a) \left( \begin{array}{c}
				x_i (\mathrm{1115})\\
				y_i (\mathrm{1115})\\
	 			F_i (\mathrm{1115})
				\end{array} \right) 
				\hspace{1 cm}
				 (0 \leq a \leq 1) 
\end{eqnarray}

which is parameterized solely by the interpolation factor $a$. We used photometry from the CASTLES web-site \citep{castles} to set the flux ratios and relative positions of the components of H1413$+$117 and PG1115$+$080, but we normalized the total flux of each lens to be the same observed median flux we used for our fiducial two-component models. Figure 10 illustrates such an interpolated sequence in a pictorial manner. In addition to this interpolation, we also allow the scale of the entire system to change, corresponding to changing the normalization of the lensing mass. Although the constant size of the pixels sets a certain scale, changing the inter-component separations can also be thought of as a rough proxy for changing the seeing, since the ratio of the separations to the FWHM changes. Figure 11 shows the regions of this two dimensional symmetry - size parameter space in which our simulated quads pass the $\Delta \chi^2 > $ 0.1. Here, a scale of one corresponds to a sequence of quads where the endpoint geometries have the same size as the real systems H1413$+$117 and PG1115$+$080. Here we again use the median observed total flux and seeing. Although our selection technique is by construction best-suited to detecting two-component objects, figure 11 illustrates that it is in fact quite sensitive to any deviations from a point-source morphology. Particularly encouraging is the recovery of 1413-type systems which, \emph{a priori}, one might expect to be problematic due to their considerable radial symmetry. 
  
\subsection{Truth Testing}

Formally, the selection function derived above only sets an upper limit on the detectibility of a particular lens configuration because our simulations do not include all possibly relevant observational effects. For instance, they do 
not include: i) PSF anisotropy. We use circularly-symmetric PSF's in our
simulations whereas the observed SDSS PSFs are often (and sometimes highly) non-symmetric. ii) Non-Poisson noise. Bright stars, faint objects, and cosmic rays can all contribute sky noise which is not represented by a Poisson distribution. Since our $\chi^2$ minimization is weighted by the square of observed pixel counts, we might expect that our fits will be more accurate in simulation where this weighting corresponds to the true variance. Even if such effects do not systematically reduce the detectibility of non-point source morphologies, they at least are an unmodeled source of noise which might statistically reduce the number of lenses detected. 

Another potential difference between our simulations and the real data is the performance of the simplex minimization scheme. We know that sometimes the simplex minimization will converge to a local, non-global, minimum. For the purposes of measuring the selection function, it is irrelevant whether or not this actually happens, only whether or not it happens more (or less) frequently for real than for simulated data. In fact, we know from the accuracy of the best-fit geometries, figures 6 \& 7, that the correct geometry, and therefore global minimum, is being found for simulated objects with $\Delta \chi^2 > 0.1$. Hence, our truth testing will tell us whether or not the global minima are being found for the real data as well.

We could test the importance of all remaining unmodeled effects if we were able to compare our simulation results to known pairs of point-sources in the SDSS dataset. One obvious approach would be to investigate how our two-component fit performs for known lenses. We performed such a comparison for the newly-discovered lens SDSS1226-0006 \citep[henceforth I02]{Inada1}. This system has two quasar images with a fairly small flux ratio between them. The purpose of this comparison is largely to give the reader a qualitative impression of the accuracy of the best-fit geometry . Table 1 shows how the geometries recovered from our fit to SDSS imaging compare to those measured from  high-resolution imaging with the Walter Baade 6.5m telescope as reported in I02. In all three SDSS bands, our fit is consistent with the higher-resolution imaging to within the errors derived in section 2.2 . Table 2 compares the measured  $\Delta \chi^2$ to a set of simulated objects of the same geometry, flux and PSF\_WIDTH. The quoted errors for the simulated values are the observed standard deviations among 16 realizations of the same object. The simulated values in $r$ and $i$ appear to be systematically somewhat higher than the real values, although this may be in part due to flux from the lensing galaxy. Unfortunately, few meaningful conclusions can be drawn on the basis of a single object. Further, the set of all previously-known lens systems imaged by the SDSS to date turns out to be an unsuitable sample for our present purposes; as will become evident when we discuss the properties of these objects as observed by the SDSS in section 3.3 .  

As an alternative, we can take advantage of a minor failing of the SDSS galaxy target selection to obtain a sample of close pairs of point sources. Since, as demonstrated in the preceding section, the PHOTO deblender generally does not deblend pairs of point sources at separations below about $3^{\prime\prime}$, binary stars at these separations will be classified by the star-galaxy separator as galaxies and, consequently, approximately $1\%$ of all objects targeted in the SDSS as galaxies are in fact spectroscopically-identified as stars \citep{scranton}. We selected such objects from the EDR and visually-inspected the atlas images to reject star/galaxy pairs and other obvious contaminants. This procedure yielded a sample of 307 objects which, although we cannot be certain of their true geometries, we believe to be at least dominated by binary stars. We then fit a two-component model to each of these objects to recover a best-fit separation, flux ratio and total flux, as well as a measure of the $\Delta \chi^2$. We also recorded the PHOTO measured PSF\_WIDTH for each object. Next, for each object we simulated a pair of point sources with the same seperation, total flux, and PSF\_WIDTH. We then fit these simulated objects with a two-component model and compared the value of the measured $\Delta \chi^2$ for the simulated object to that of the real object upon which it was modeled. Figure 12 illustrates our truth testing procedure as a flowchart. Figure 13 shows the distribution of fractional differences between the simulated and observed value, 

\begin{eqnarray} 
	f(\Delta \chi^2) \equiv \frac{\Delta \chi^2_{observed} - \Delta \chi^2_{simulated}}{\Delta \chi^2_{observed}}
\end{eqnarray} 

 If the simulated $\Delta \chi^2$ are systematically higher (or lower) than the real values, then the center of this distribution should offset from zero. Estimating the center is somewhat complicated by the fair number of outliers, particularly for $f(\Delta \chi^2$) $<$ -1 (the bins at either end of the histogram contain all points at or beyond those values). We attribute these outliers mainly to objects whose underlying geometry is not in reality a pair of point sources. Clearly in such a case the $\Delta \chi^2$ for the simulated best-fit two-component geometry would be expected to be greater than the observed value. However, the goal of this section is to get an estimate of the systematic error in $\Delta \chi^2$ for cases in which the underlying geometry really is a pair of point sources, hence ideally we would like an estimate which ignores such outliers. One way to do this is simply to take the median value, which is $f(\Delta \chi^2$) = 0.02 . Another is to fit a Gaussian to the distribution at $-1 < f(\Delta \chi^2) < 1$, as shown in figure 13. In this case, we find the center at $f(\Delta \chi^2$) = 0.07 . In either case, the systematic error in $\Delta \chi^2$ is $< 10$ \%, and in fact indicates that the simulations somewhat under-predict the value of $\Delta \chi^2$. Ideally, we could also compare the dispersion of this distribution to the dispersion of a similiar simulated distribution to estimate the random error introduced by unmodelled properties of the data. Unfortunately, the objects in this distribution are taken from the main galaxy sample and hence are considerably brighter than the vast majority of the quasars in which we are most interested. As a result, for instance, the effect of correctly modelling the sky noise should be much less important for these objects. Hence, at this time, we satisfy outselves with our conclusion that the systematic differences between the real data and our simulations are small ($< 10$ \%) and that, consequently,  the selection functions derived in the previous sections can be regarded as having been calibrated with real SDSS data. 

\section{Application to SDSS Data}

\subsection{Observational Details}

	The SDSS is a photometric and spectroscopic survey across 10,000 square degrees of the northern Galactic cap using the 2.5m SDSS telescope at Apache Point Observatory. SDSS imaging is carried out with a wide-field camera \citep{1998AJ....116.3040G} with makes nearly simulataneous observations of objects in five passbands: {$u$ $g$ $r$ $i$ $z$}. Together, the passbands cover the optical wavelengths from the atmospheric cut-off in the blue to the minimum detectible energy for the silicion CCD's in the red \citep{1996AJ....111.1748F}. Photometric calibration of the imaging survey is seperately carried out by an automated 0.5m telescope \citep{2001AJ....122.2129H} which monitors a set of standard stars \citep{2002AJ....123.2121S} while photometric data is being acquired. The SDSS imaging camera also incorporates astrometric CCDs which provide astrometry of detected objects with an acccuracy typically better than 0$\farcs1$ \citep{Pier}.  

\subsection{Selecting Lens Candidates}

	In this section we apply the selection techniques we described and analyzed in section 2 to a sample of SDSS quasars. We define as our parent sample the 5120 quasars from the EDR imaging which are confirmed by SDSS spectroscopy to be at $z$ $>$ 0.6. We chose to restrict ourselves to this sub-sample of the imaging data because at present we are primarily interested in presenting methods and the EDR represents a well-vetted test-bed. Our parent sample is larger than the 3814 quasars of the SDSS EDR quasar catalog \citep{2002AJ....123..567S} primarily because we do not restrict ourselves to objects whose spectra appear on EDR spectroscopic plates. The SDSS chooses sets of spectroscopic targets (plates) from the imaging data (runs). However, not all of the plates targeted from EDR imaging runs are included among the plates released as part of the EDR. Hence, our sample additionally contains objects whose spectra were targeted on EDR imaging data but not released as part of the EDR. Although this may at first seem a somewhat confusing selection, it arises naturally from the fact that we defined our sample solely in terms of the $\sim$ 400 sq. deg. of EDR imaging data and then simply found all objects which had SDSS spectroscopy, without regard to whether or not those spectra were part of the EDR or not. Another, more minor, difference between our sample and the SDSS EDR quasar catalog is that we do not exclude any objects based on visual inspections of the spectra or requirements on equivalent widths of emission lines.   
	To define a sample of lens candidates, we sequentially apply four cuts to our parent sample. Below, we present these cuts together with the number of objects which remain after each cut and further comments as appropriate. 

i) Morphological $\Delta \chi^2$ Cut: We require that the (S/N)$^2$-weighted average of $\Delta \chi^2$ as fit in $g$,$r$, and $i$ be greater than 0.1. 408 objects remain after this cut. 

ii) Separation Cut: We require that that the average best-fit separation as fit in $g$,$r$, and $i$ be less than 3$\farcs$0. 105 object remain after this cut. 

	As discussed in section 2.1, we concentrated our efforts on recovering lenses with splittings $< 3^{\prime\prime}$, simply because the vast majority of lenses are expected to have such small splittings. Another reason to concentrate on small separation candidates is that while the lensing probability falls rapidly with increasing separation, the number of false positives due to coincidental alignments rises as the separation squared. \citet{1993ApJ...417..438K} calculated in detail the expected false positive rate for both stars and galaxies, but here we present the following simple estimate: From the SDSS number counts of \citet{2001AJ....122.1104Y}, we find that at the magnitudes which interest us most (19 $< r <$ 21), the number density of objects on the sky is of order 5000 per square deg. Consequently, $\sim$ 1\% of quasars will have a coincidental companion within 3$^{\prime\prime}$, a rate which is almost certainly higher than the lensing rate. Fortunately, such false positives are less problematic with SDSS data because of the colour information available from the five band SDSS photometry (see below), but without a seperation cut the efficiency of any selection is certain to be poor.

	The subject of wide-separation candidates requires an understanding of how PHOTO deals with multiple objects which are nearby one another on the sky. Here we present a brief explanation of parents and children in the SDSS catalogs. For a more complete discussion, the reader is referred to Lupton (in preparation). When PHOTO identifies an object, it includes an area \footnote{This area is the predicted full extent of the object based upon the empirical PSF} of the sky around the object in the atlas image. If another object appears within that area, then the two (or more) objects are listed together in the SDSS catalogs as a single compound object (parent). Where appropriate, PHOTO then attempts to deblend the parent into single objects (children) and, when successful, the deblended children also appear in the SDSS catalogs. Since a parent object can contain children separated by many arcseconds, a large fraction of quasars will have wide-separation parents. SDSS quasar target selection only considers the children of deblended parents, but we fit both the children and their parent (if any). Consider, for example, a deblended pair of quasars $2^{\prime\prime}$ apart; neither child is a good lens candidate but obviously the parent would be. Conversely, a non-deblended pair of quasars $1^{\prime\prime}$ apart may have a star nearby which enters together with the pair into a parent object. In this case, both the child (close quasar pair) and the parent (wide quasar(s) plus star pair) might pass our morphological selection but only the close pair would pass our separation cut.  

iii) Consistent Geometry Cut: We require that $\Delta \mathrm{position} < $ 0.4 arcseconds geometric consistency between the best-fit models in $g$,$r$, and $i$ as defined in eqn. 12 . 41 objects remain after this cut

iv) Colour Cut: We require that the colour differences between best-fit model components be consistent with a simple reddening model. Specifically, we require that the measured $D(g-r)$ and $D(r-i)$ fall within the region defined by the inequalities

\begin{eqnarray} 
	\begin{array}{l}
	D(r-i) < D(g-r) + 0.3 \\ 
 	D(r-i) > -D(g-r) - 0.3 \\
	D(r-i) > \frac{7}{9}D(g-r) - 0.3 \\ 
	D(r-i) < - D(g-r) + 1.3 
	\end{array}
\end{eqnarray}

	Note that in order to avoid ambiguities of sign we define colour differences such that $D(g-r) > 0$ and then measure $D(r-i)$ relative to the bluer component in $g-r$. 13 objects remain after this cut. 

We base this cut on the Figure 1 of F99 which indicates that, apart from a pair of very highly-reddened systems, the distribution of observed differential extinction for known lenses goes to zero at a value of $\Delta E(B-V) \simeq 0.4$. We devised a simple model to predict the colour difference between two quasar images due to reddening in the lens galaxy. In what follows, we assume that one of the quasar images is unreddened, so that the colour difference is identical to the reddening of the other image. In order to convert between $B-V$ and SDSS colours, we need a quasar SED and a reddening law. For redshift $\leq$ 2, we use the composite quasar spectrum of \citet{2001AJ....122..549V}. Beyond redshift 2, Lyman-alpha enters the B-band and this composite does not correctly predict quasar colours due to Lyman forest absorption. Instead, at redshift 3, we constructed our own composite using the spectra of 198 SDSS quasars with redshifts $2.5 < z < 3.5$ . Following F99, we use the Galactic reddening law of Cardelli, Clayton, \& Mathis (1989) \nocite{1989ApJ...345..245C}. We chose a value for the quasar redshift, reddened the composite spectrum as appropriate for a chosen lens galaxy, and then convolved the reddened spectrum with the filter transmission curves to find the observed colours. By doing this for a number of values of the total extinction, $A_V$, we can find the values of $D(g-r)$ and $D(r-i)$ which correspond to different values of $D(B-V)$ and thus construct reddening vectors in the $D(g-r)$ - $D(r-i)$ colour space. Figure 14 shows nine reddening vectors of length $D(B-V) = 0.5$ for quasar redshifts $z_{qso}$ $\in$ \{1 2 3\} and lens galaxy redshifts $z_g$ $\in$ \{0.3 0.5 0.7\}. The inclusion region described in eqn. 16 is also shown. We assume reddening in our galaxy has a constant value of $A_{V,galactic} = 0.1$ and that $R_V$ = 3.1 . F99 found some evidence of differing values of $R_V$, but presently we do not consider this possibility in our simple model. Simply stated, this cut requires that if a colour difference is observed in $g-r$, then the observed colour difference in $r-i$ must be consistent with this simple reddening model and that the observed total differential reddening be such that $D(B-V) < 0.5$

Table 3 summarizes the results of this four-step selection.  
	
	Of the remaining 13 objects, 6 have received follow-up observations which can with some certainty confirm or rule out the lensing hypothesis. One of the objects is SDSS1226-0006, and five are not lenses. The objects which have been followed-up cannot be taken as a representative sample because follow-up efforts tend to favour the brighter, wider separation, and more equal flux ratio candidates, but if we naively ignore this bias, then we observe a lensing rate of 2 out of 5120, or 4 $\times 10^{-4}$. If the lensing probability follows a Poisson distribution, then our sample of 13 candidates implies an upper bound on the observed lensing fraction of not more than 22 out of 5120, or 4 $\times 10^{-3}$ and the one lens in our sample implies a lower bound of not less than 3 $\times 10^{-5}$ (95\% confidence levels).  

\subsection{Previously-Known Lenses}

An obvious question to ask regarding any selection technique is how well it recovers previously-known lensed quasars. If very few previously-known systems are recovered, then it stands to reason that some the assumptions we have made in designing our selection technique were incorrect. On the other hand, we must guard against overstating the importance of recovering known systems. The sample of previously-known systems is very heterogenous; subject to a variety of, often unknown, selection effects. Hence, if we were to overly modify our selection specifically to recover previously-known systems, we might well introduce those same selection effects into our sample.

At the time of writing, twelve CASTLES lens/binary systems have been imaged by the SDSS. In addition, the current survey area also includes two newly-discovered lens systems, SDSS1226-0006, and SDSS0924+0219 \citep{Inada2}. Both of these systems were initially identified by independant means. Hence, a total of fourteen previously-known systems have been imaged at present. Of these, five have image separations greater than $6^{\prime\prime}$. Such systems are deblended by PHOTO and not in the regime of interest for this work. A further three are much too faint (1 magnitude of more below the $i$ = 20.2 flux limit of the high-redshift quasar sample) for SDSS spectroscopy and hence out of our parent sample. There follows a discussion of the results of our selection for each of the remaining six systems: 

\textbf{APM08279+5255} : \citep{1998ApJ...505..529I} This is a very bright three-image lens with two image of roughly equal brightness separated by 0$\farcs$38 and a third fainter image (flux ratio $\sim$ 11:2) roughly halfway between them \citep{2002MNRAS.334L...7L}. This object is actually too bright for SDSS spectroscopy and hence is not targeted as a quasar. We recover the correct separation and flux ratio of the two brighter components in $r$ and $i$, but not in $g$ where the object is almost two magnitudes fainter ($g$,$r$,$i$ = 17.2, 15.4, 14.9) and the seeing is worse. Consequently, it passes our $\Delta \chi^2$ and consistent geometry cuts but fails the colour cut due to a large, best-fit, $g-r$ colour difference. Given the image seperation, our ability to recover the geometry at all is quite encouraging, albeit at extremely high signal-to-noise. 

\textbf{SBS0909+532} : \citep{1997ApJ...479..678K} This is a bright two-image lens with an image separation of 1$\farcs$1. It is targeted by SDSS quasar target selection and passes all of our lens candidate selection criteria. 

\textbf{SDSS0924+0219} : This unusual system appears to have only three bright quasar images, but is in fact almost certainly a four image system with the fourth image obscured and/or demagnified by micro-lensing. It is targeted by SDSS quasar target selection and passes all of our lens candidate selection criteria. This object demonstrates that our method can recover lenses with more complicated geometries than a simple two-image configuration. 

\textbf{SDSS1226-0006} : This is a two-image lens  with an image separation of 1$\farcs$2. It is targeted by SDSS quasar target selection and passes all of our lens candidate selection criteria. 

\textbf{SBS1520+530} : \citep{1997A&A...318L..67C} This system has a rather complex morphology (there are up to four quasar images and up to three lens galaxies). Furthermore, it is within 14$^{\prime\prime}$ of a bright ($r$ = 13.6) star. Presumably due to being in the PSF wing of this star, the object is badly-deblended into four children of non-astrophysical colours. Consequently, the results of our fitting to this object are not meaningful. The object is also not targeted by SDSS quasar target selection.

\textbf{HST15433+5352} : \citep{1999AJ....117.2010R} This is a faint two-image lens system with an image separation of 1$\farcs$2. The lensing nature of this object is still not certain. It is slightly too faint for SDSS quasar target selection ($i$ = 20.6). Additionally, it is classified by the PHOTO star-galaxy separator as a galaxy, and hence, it would not be targeted as a high-redshift quasar due to its extended morphology. This is an example of the anti-lens bias of the high-redshift quasar sample which was discussed in section 2.3.1 . This object passes our $\Delta \chi^2$ and consistent geometry cuts but fails the colour cut due to a large best-fit $r-i$ colour difference. This colour difference is due to a somewhat inaccurate fit in $r$ which is caused by the lower signal-to-noise in this band ($r$ = 21.6). 

The results of our selection for these six objects are summarized in table 4. For the three objects which we do select, our selection techniques work just as envisioned; we obtain a robust $\Delta \chi^2$ in all three fitted bands and the measured colour differences are small. For the two objects which are not selected, inaccuracies in the best-fit model in one or more bands lead to large colour differences which cause the object to be rejected as a lens candidate. These two examples show that our colour cuts are in fact also cuts on the accuracy of the two-component model when the object lies near the selection threshold. However, given that one of these two failures occurs at a very small splitting angle and the other at a signal-to-noise below the limit of the SDSS quasar sample, we feel that neither warrants changing our selection criteria. That said, as more known systems are imaged, we intend to repeat this exercise and re-appraise our selection techniques accordingly.

\section{Discussion}

	The main goal of this work has been to present our method for quantifying the selection function of our (or any) selection algorithm. We have explicitly presented the selection function only for a somewhat coarse sampling of the available parameter space, particularly for four-component systems, but our simulation method can be straight-fowardly applied to determine the selectibility of any lens geometry. The accuracy to which the selection can be determined is limited by the behaviour of unsimulated properties on the real data, as considered in section 2.4 . It ceases to be useful to further improve one's understanding of the selection function once the error associated with it is much less than the uncertainty introduced in the statistics of the lens sample by Poisson fluctuations. If the lensing rate is indeed $< 10^{-3}$, then the Poisson uncertainty for the complete sample of $10^5$ SDSS spectroscopic quasars will be $>$ 10 \%. Such an accurary should be readily-accessible through our methods, at least for the morphological selection.
 
	In order to make use of our selection function, one needs to predict the distribution of lens geometries so as to correct the observed distribution for those lenses which will not be selected. We have not undertaken such an exercise at present, consequently our upper bound on the observed lensing fraction is only a lower bound on the upper bound on the true lensing fraction. Presently, we feel that attempting to compute the selection function for the whole survey would be premature. There are still a few obvious improvements to our selection procedure, such as an improved seeding scheme to recover wider separation candidates, which should be made. Also, as stated above, the data, in particular the seeing conditions, of the EDR are probably not representative of the whole survey. Additionally, since the prediction of lensing statistics requires knowledge of the quasar luminosity function, it is sensible to await determination of the SDSS luminosity function since, almost by definition, it will be luminosity function best-suited to the this aim.

	A theoretical model of the lens galaxy population will also be required to better constrain the effects of the lens galaxy flux upon the selection function. The estimates in section 2.1 imply that our simplifying assumption that the lens galaxy flux is negligible probably introduces moderate errors in the selection function for the majority of quasars at redshifts less than 3, and significant ones at higher redshifts. A more detailed calculation of this effect is required. If, in fact, many lenses have prominent lens galaxies, then our simulation techniques could still be used to measure the selection function, but the accuracy of the result would be dependant on the accuracy with which we could predict the photometric properties of the lens galaxies.

	Another unresolved problem is the use of colour as a selection criterion. On the one hand, to not use the multiple colours available from SDSS imaging is a enormous waste of information, particularly when attempting to reject stellar interlopers. For that matter, we have been conservative in this work by not using $u$-band imaging although, for quasars with $z < 2.5$, $u-g$ colour is the strongest discriminant between quasars and stars. On the other hand, while our simulations allow for an accurate quantification of the selection function associated with our morphological criterion, the selection effects associated with colour selection remain difficult to similarly characterize. Further, if, contrary to our expectations, a sizable fraction of lenses are in fact being reddened, or otherwise displaced in colour space, so as to fall onto the stellar locus, and hence out of the quasar sample, then there is little prospect of ever accurately measuring the lensing rate. 

	There are a number of obvious observational extensions to the work presented in this paper. First would be to complete the follow-up of the candidates identified in section 3 so as to get an actual measure of the observed lensing rate for this sample. Beyond that, applying our selection techniques to the full set of available SDSS quasars (there are approximately 30000 at the time of writing) would produce a sample of candidates which should give a more authoritative estimate of the lensing rate. A sample of this size, likely several hundred candidates, would probably best be approached by beginning with a program of high-resolution imaging to confirm candidate geometries before attempting the telescope-intensive high spatial-resolution spectroscopy or near-IR imaging which is required to definitively test the lensing hypothesis.   

	As shown in section 2.3, many of the most prominent lenses at redshifts greater than 3 will not be in the SDSS quasars sample due to their extended morphology. However, due to higher lensing rate at these redshifts, it may be profitable to attempt to identify such lenses from the SDSS imaging alone. In particular, if our morphological criterion can be refined to allow for the selection of lens geometries from among normal, red galaxy morphologies, then it may be possible to carry out an efficient survey for lenses at these redshifts. 
	
 	We draw attention to the possibility of the application of our simulation techniques to problems beyond the discovery of gravitational lenses. Other researchers have previously used simulated SDSS images to study selection effects (cf. Kim et al. 2002\nocite{2002AJ....123...20K}), but the simulation procedure we have described has the advantage of allowing one to create simulated images of any sort of object with realistic noise properties, particularly when the simulations are calibrated by real data as in section 2.4 . Such calibrated images are well-suited for any undertaking which involves additional processing of the SDSS imaging data. Bulge-disk decomposition and other measures of galaxy shapes beyond the measured PHOTO parameters are obvious examples of problems which might be approached in this way. Additionally, because the simulations are passed through the same reductions as the real data, one has the ability to investigate how a given class of objects is treated by PHOTO. One possibly interesting application of this would be to investigate the treatment of bright galaxies by the PHOTO deblender. 

Comparison of our preliminary observed lensing fraction with previous results is complicated by the absence of a comparable homogeneous sample. The HST Lens Snapshot Survey \citep{1993ApJ...409...28M} found 6 lenses among 502 optically-selected quasars (a rate of 1.2 $\times 10^{-2}$), but the magnification bias is likely much more pronounced for their brighter parent sample. The CLASS survey found 22 lenses among 11,685 flat-spectrum radio sources (a rate of 1.9$\times 10^{-3}$), but in this case the uncertain redshift distribution and optical luminosity function of their sources again makes direct comparison difficult. A definitive measurement of the lensing rate from the SDSS will not only await further survey data, but also a great deal of observational follow-up outside of the scope of the survey itself. 

\acknowledgements

We thank Gordon Richards and Michael Strauss for a number of useful discussions. This work was supported by NASA grant NAG5-9274. 

Funding for the creation and distribution of the SDSS Archive has been provided by the Alfred P. Sloan Foundation, the Participating Institutions, the National Aeronautics and Space Administration, the National Science Foundation, the U.S. Department of Energy, the Japanese Monbukagakusho, and the Max Planck Society. The SDSS Web site is http://www.sdss.org/. 

The SDSS is managed by the Astrophysical Research Consortium (ARC) for the Participating Institutions. The Participating Institutions are The University of Chicago, Fermilab, the Institute for Advanced Study, the Japan Participation Group, The Johns Hopkins University, Los Alamos National Laboratory, the Max-Planck-Institute for Astronomy (MPIA), the Max-Planck-Institute for Astrophysics (MPA), New Mexico State University, University of Pittsburgh, Princeton University, the United States Naval Observatory, and the University of Washington.
  
\appendix

\section{Simulating SDSS Images}

In this appendix, we describe how we created the simulated SDSS images which we used in section 2 to measure the selection function of our morphological selection criterion. We begin by choosing a value for the PSF FWHM and creating a simulated image of a field of bright point sources with that FWHM. The image is generated by first integrating analytic PSFs on a grid corresponding to the SDSS pixel scale of 0$\farcs$4 /pixel. We used an analytic PSF composed of two Moffat functions with $\beta_1 = 7$ and $\beta_2 = 2$, as recommend by \citet{1996PASP..108..699R}. It should be noted that PHOTO uses a double Gaussian model when estimating the PSF\_WIDTH in a frame and, consequently, we find that the measured PSF\_WIDTH is systematically greater than the input FWHM of our Moffat functions. To first order, the relation between the two is PSF\_WIDTH $\simeq$ 1.15 $\times$ FHWM$_{\mathrm{Moffat}}$. Throughout the previous sections of this paper quoted FWHM values have been for the PSF\_WIDTH as reported by PHOTO. We then add a constant sky level and assign 'observed' counts for each pixel as a random Poisson variable with a mean value of the object plus sky counts. For the PSF fields we used the observed median EDR sky value in $r$. We also add a Poisson random variable corresponding to the observed dark variance. Having recorded our field of PSF stars as a standard FITS file, we then pass this simulated field through to the so-called fake stamp collection pipeline (fsc\_pipeline) which produces ancillary files required for running PHOTO. The image is then passed through the postage stamp pipeline (ps\_pipeline) which outputs an empirical PSF (psField). It should be noted that, in contrast to the PSF\_WIDTH estimate, this empirical PSF does not assume a double Gaussian or any other particular functional form (again, see Lupton et al. 2001\nocite{2001adass..10..269L} for details). The procedure for simulated lens candidates is the same, except that instead of single point sources we lay down a pair of point sources with the desired separation, flux ratio, and signal-to-noise. Also, for simulated ensembles where each object is reproduced in ten iterations, we folded-in the observed distribution of sky values by assigning to the first iteration, the 5th-percentile sky level; to the 2nd iteration, the 15th-percentile sky level,
 and so on. The sky distribution we used was simply the observed sky values in the 5120 fields which contain our SDSS quasar sample. Again, we prepare to run PHOTO (fsc\_pipeline) and then we pass our image through the frames pipeline (frames\_pipeline) which, making use of the empirical PSF which we just previously constructed, outputs the cropped atlas images (fpAtlas) as well as the measured PHOTO outputs, including the results of the deblending routine. In regular survey operations, the objects and PSF field stars are in the same frame, but we perform these steps separately for efficiency as the PSF need only be measured once per choice of the FWHM. Note that we do not complete the standard photometric pipeline by calibrating the objects (nfcalib) as photometric calibration is not of interest to us. Figure 15 illustrates the simulation procedure as a flowchart. The flowchart also includes our psf-fitting to the atlas images.

\pagebreak

\pagebreak

\begin{figure}

\plotone{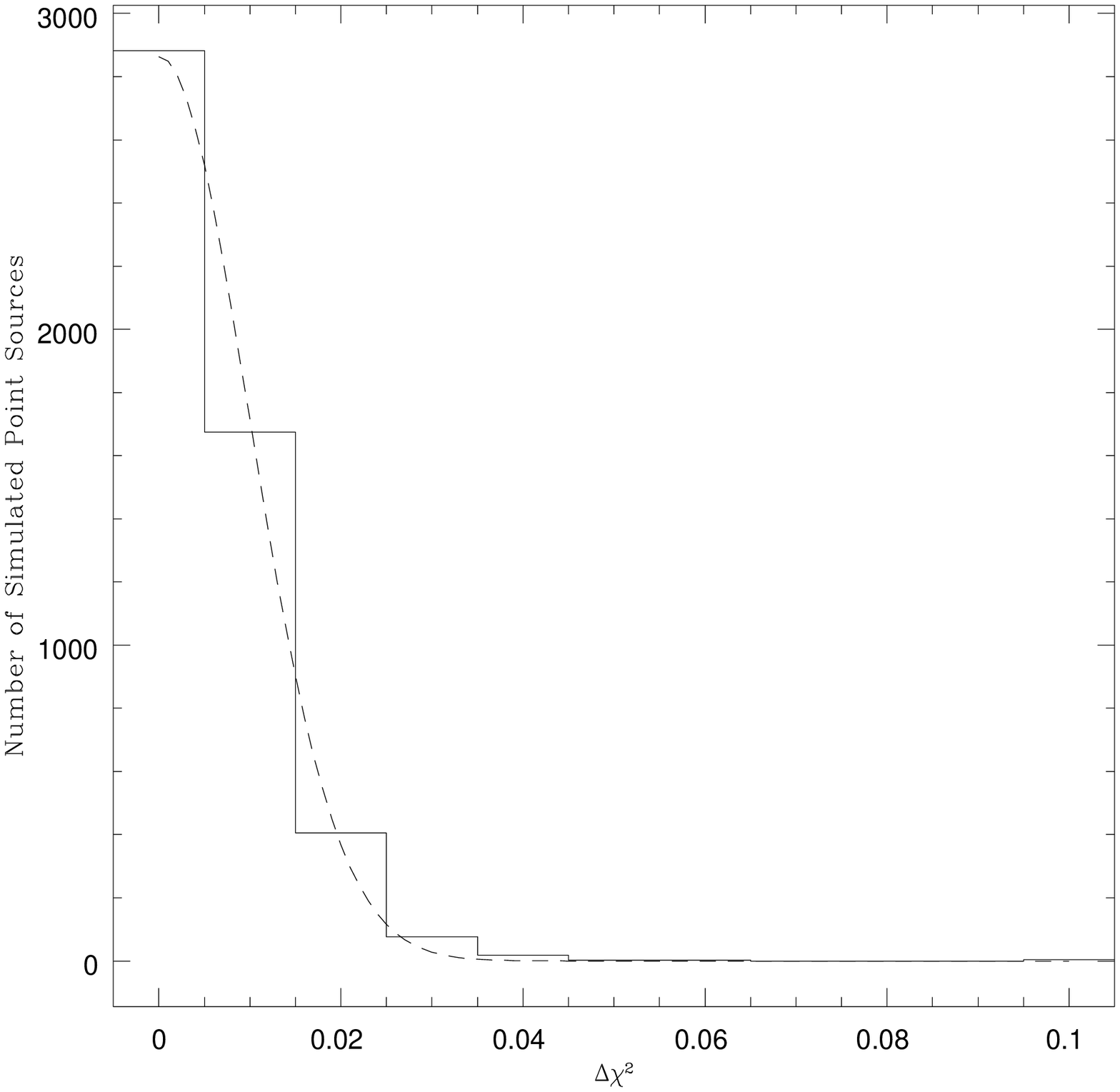}
	\caption{Solid histogram shows the distribution $\Delta \chi^2$ for 5120 simulated point sources with flux, seeing, and noise chosen to match the sample of SDSS EDR quasars. The dashed line is a Gaussian-fit to this distribution used to estimate the false positive rate as a function of the $\Delta \chi^2$ threshold.}

\end{figure}

\begin{figure}

	\centering\epsfig{figure=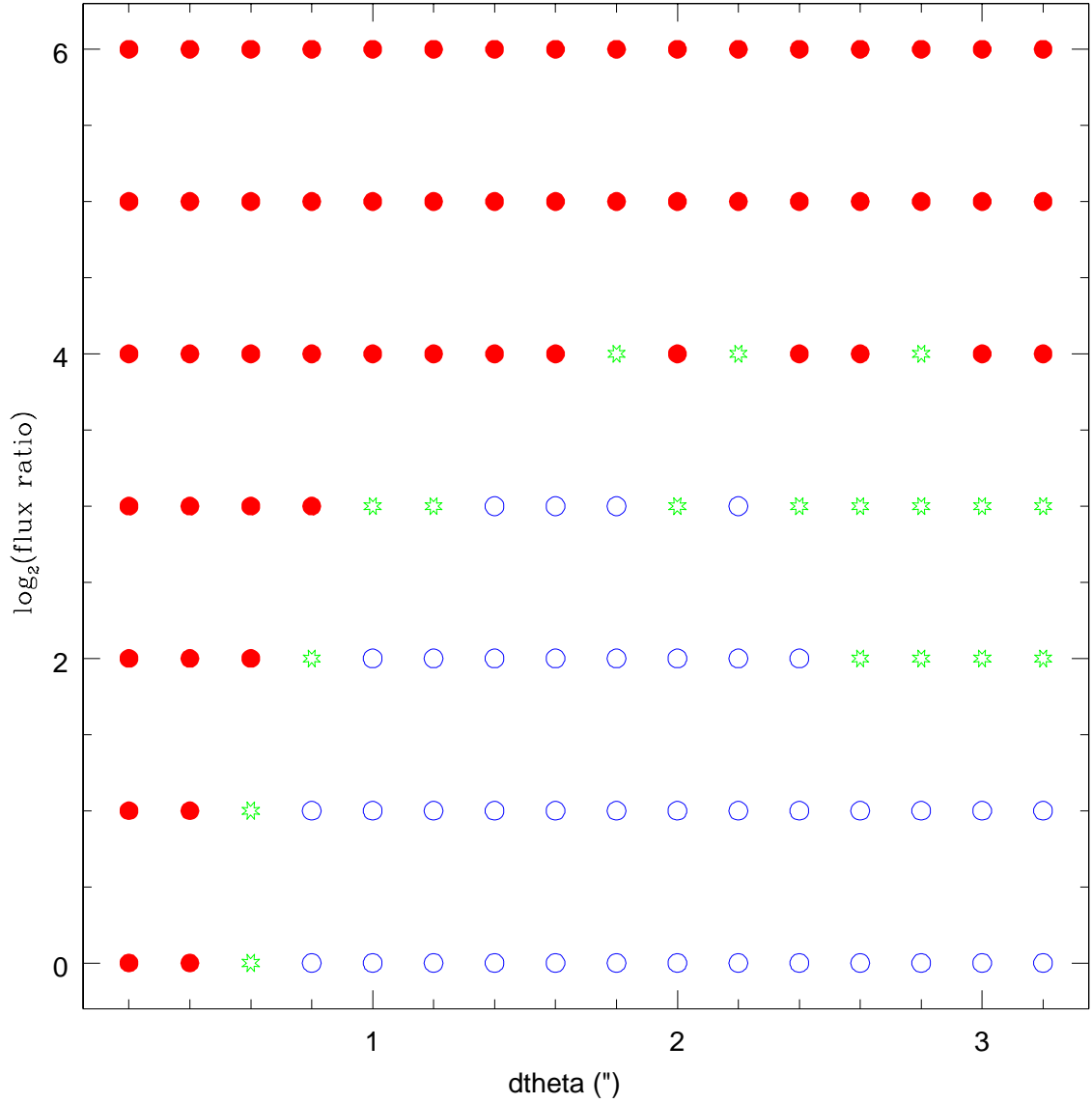,width=\textwidth}
	\caption{Regions of the separation - flux ratio parameter space in which the morphological $\Delta \chi^2 > 0.1 $ cut recovers pairs of point sources for the median seeing and flux of SDSS EDR quasars. Solid red circles indicate points where 0 out of 10 iterations are recovered. Open green stars indicate points in which between 1 and 9 out of 10 iterations are recovered. Open blue circles indicate points in which 10 out of 10 iterations are recovered.}

\end{figure}

\begin{figure}

	\centering\epsfig{figure=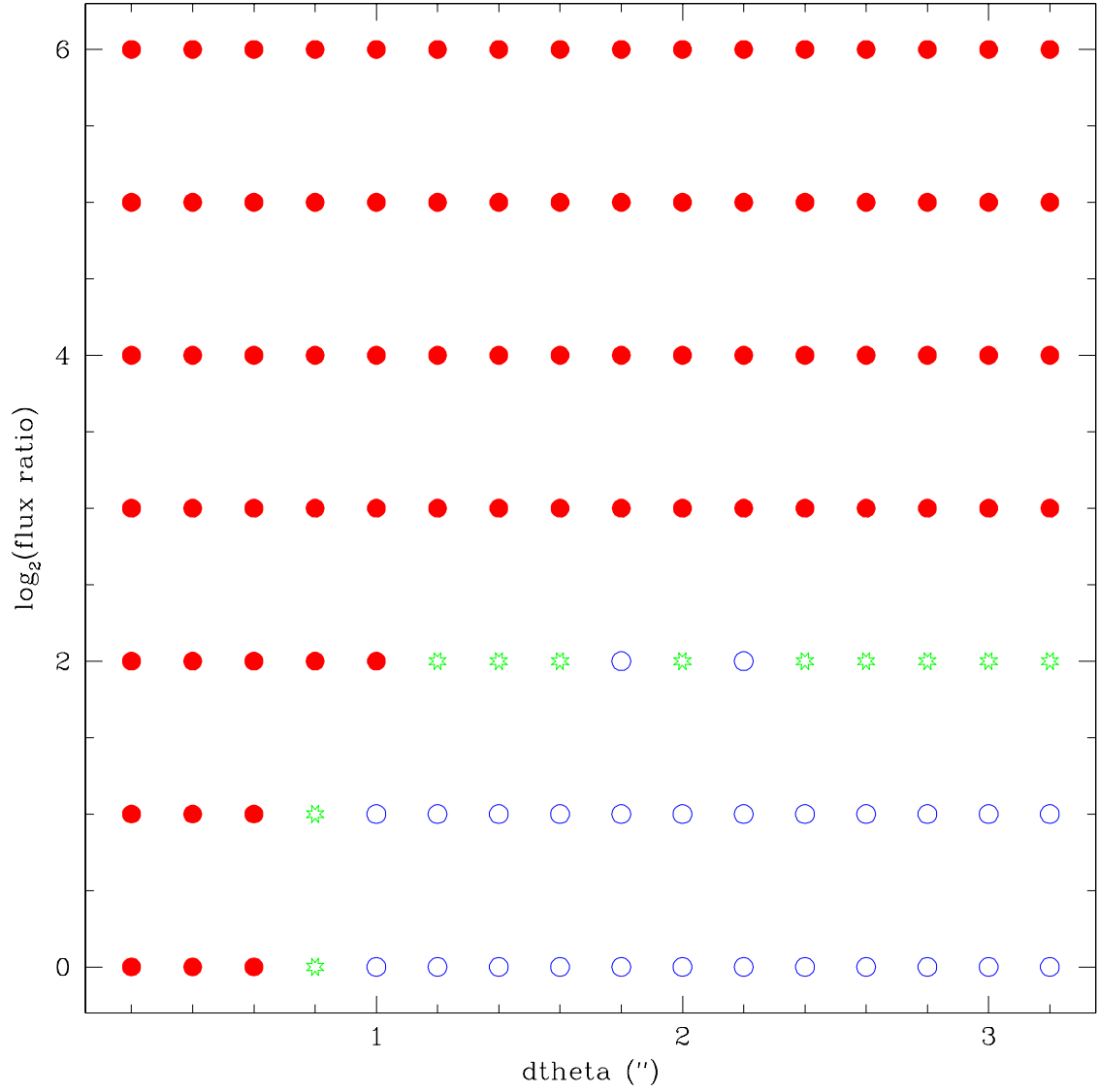,width=\textwidth}
	\caption{Same as Figure 2 except that the PHOTO star-galaxy separation is used as a selection criterion instead of the $\Delta \chi^2$ statistic. Selected points (open circles and stars) are geometries classified as galaxies by PHOTO}

\end{figure}

\begin{figure}

	\centering\epsfig{figure=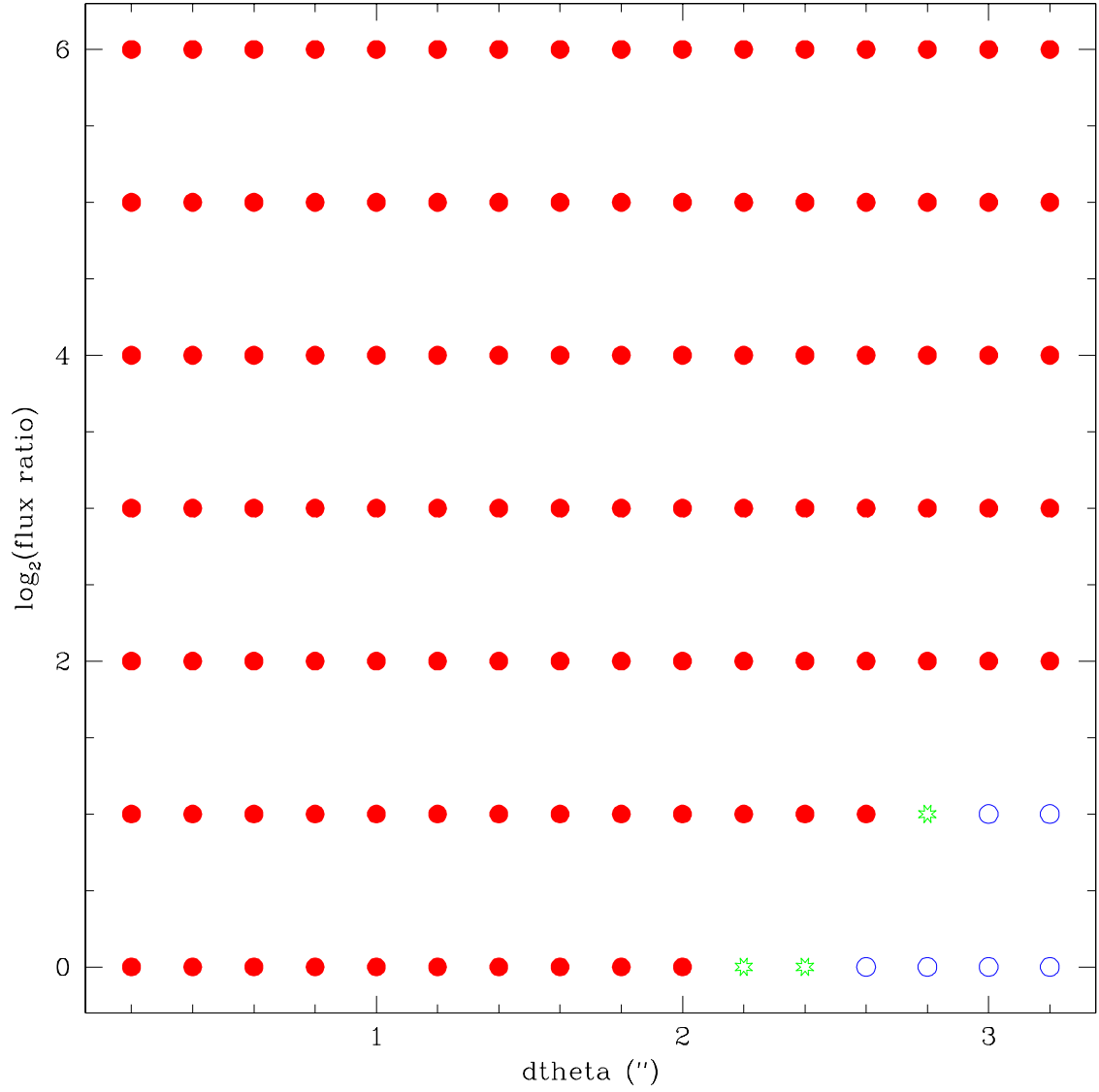,width=\textwidth}
	\caption{Same as Figure 2 except that the PHOTO deblender is used as a selection criterion instead of the $\Delta \chi^2$ statistic. Selected points (open circles and stars) are geometries deblended by PHOTO}

\end{figure}

\begin{figure}

	\centering\epsfig{figure=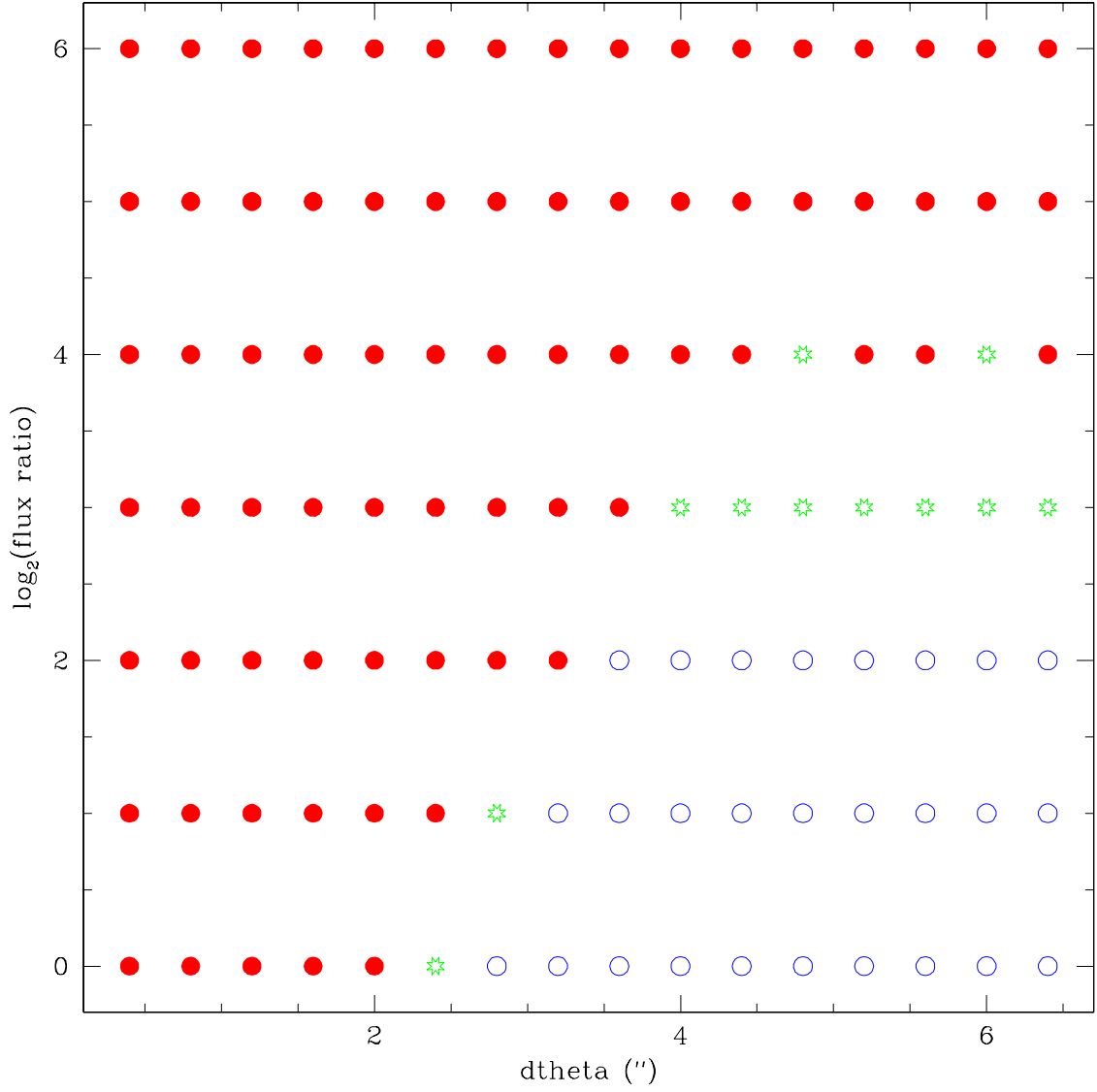,width=\textwidth}
	\caption{Same as Figure 4 except that the x-axis (separation) is expanded by a factor of two.}

\end{figure}

\begin{figure}

	\centering\epsfig{figure=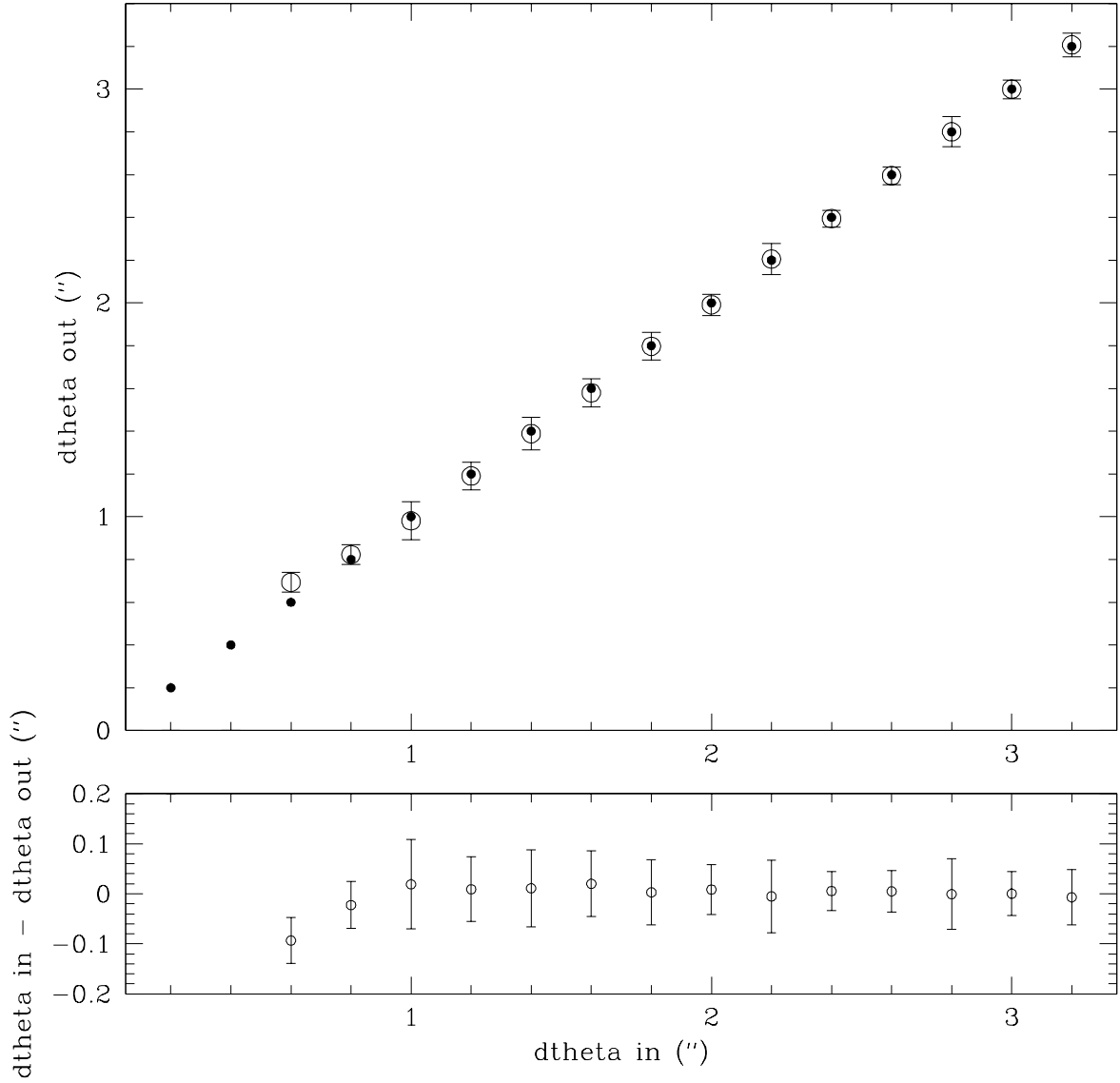,width=\textwidth}
	\caption{Upper panel shows the recovered separation of simulated pairs of point sources as a function of the input separation. Solid circles indicate the input separation. The open circles indicate the average recovered separation of all objects at a given input separation which have $\Delta \chi^2 > 0.1$ . The error bars are the measured standard deviation. Lower panel shows the difference between the input and recovered separations. }

\end{figure}

\begin{figure}

	\centering\epsfig{figure=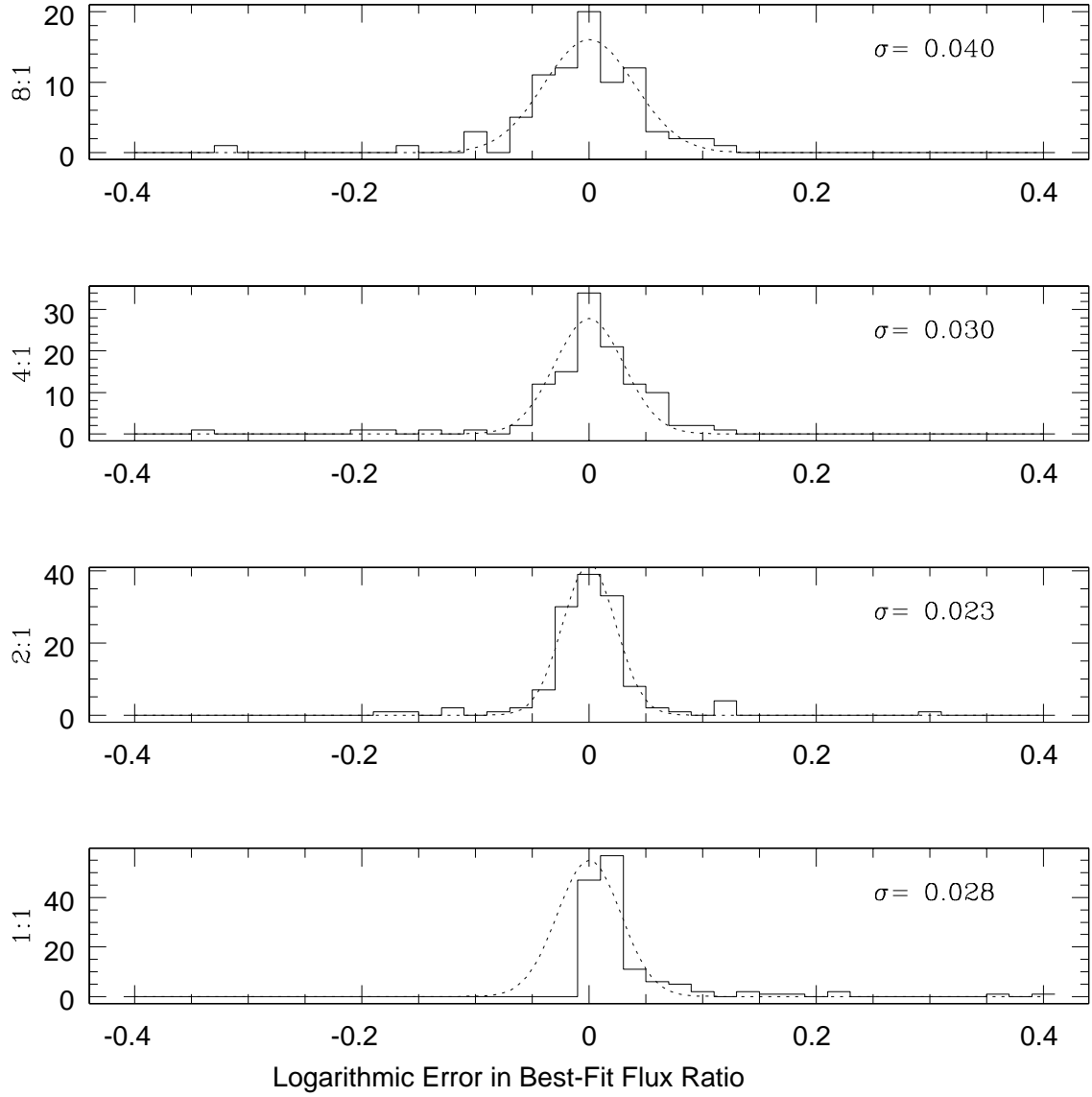,width=\textwidth}
	\caption{Distribution of logarithmic errors in the recovered flux ratio for all simulated pairs of point sources which have $\Delta \chi^2 > 0.1$ . The four panels show the distributions for different values of the input flux ratio as labeled on the y-axis. The dashed curve shows a Gaussian fit to each of these distributions. The $\sigma$ for each Gaussian is also indicated.}

\end{figure}

\begin{figure}

	\centering\epsfig{figure=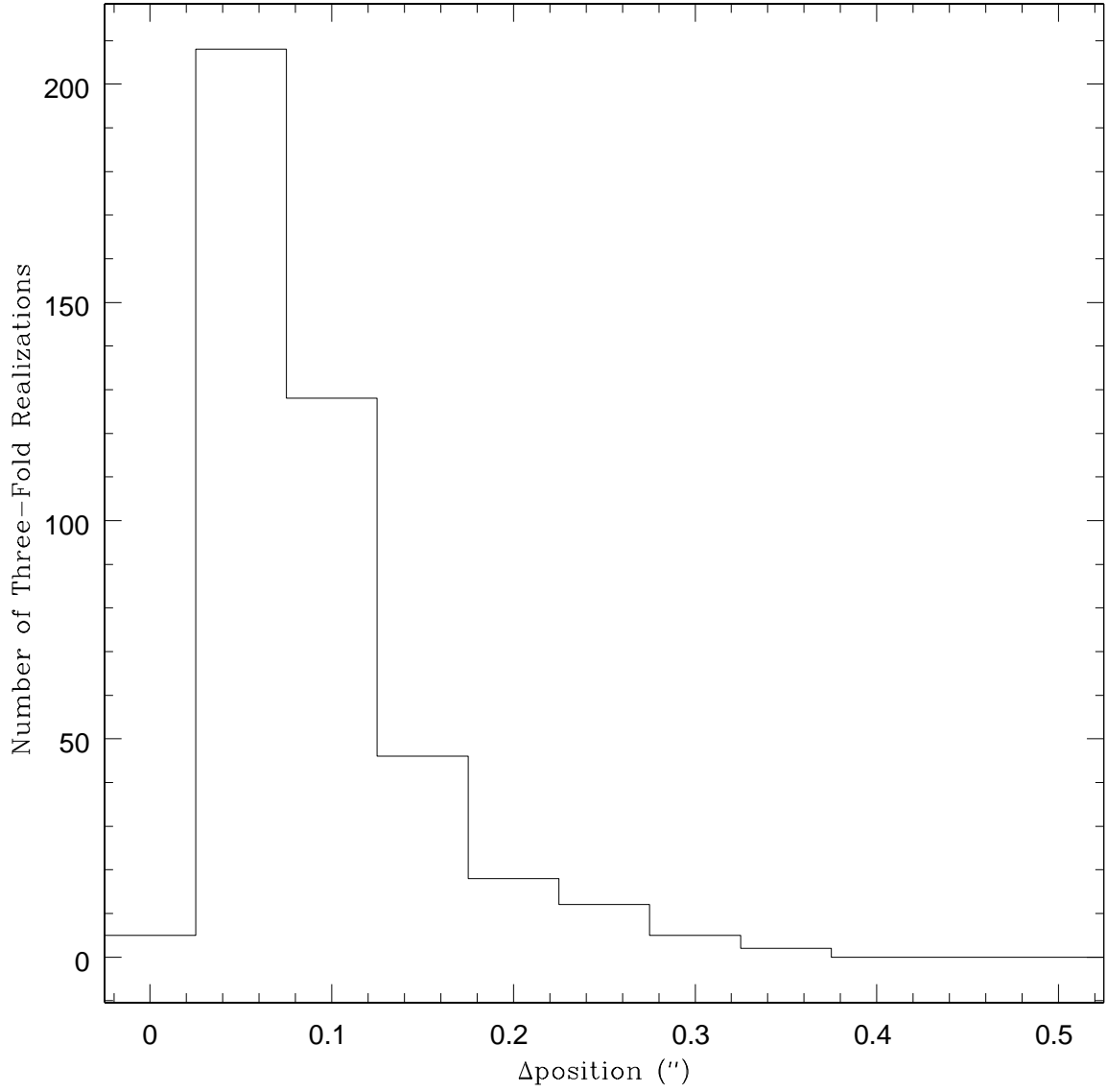,width=\textwidth}
	\caption{Distribution of $\Delta \mathrm{position}$ for three-fold realizations of pairs of simulated point sources where all three realizations have $\Delta \chi^2 > $ 0.1 .}

\end{figure}

\begin{figure}

	\centering\epsfig{figure=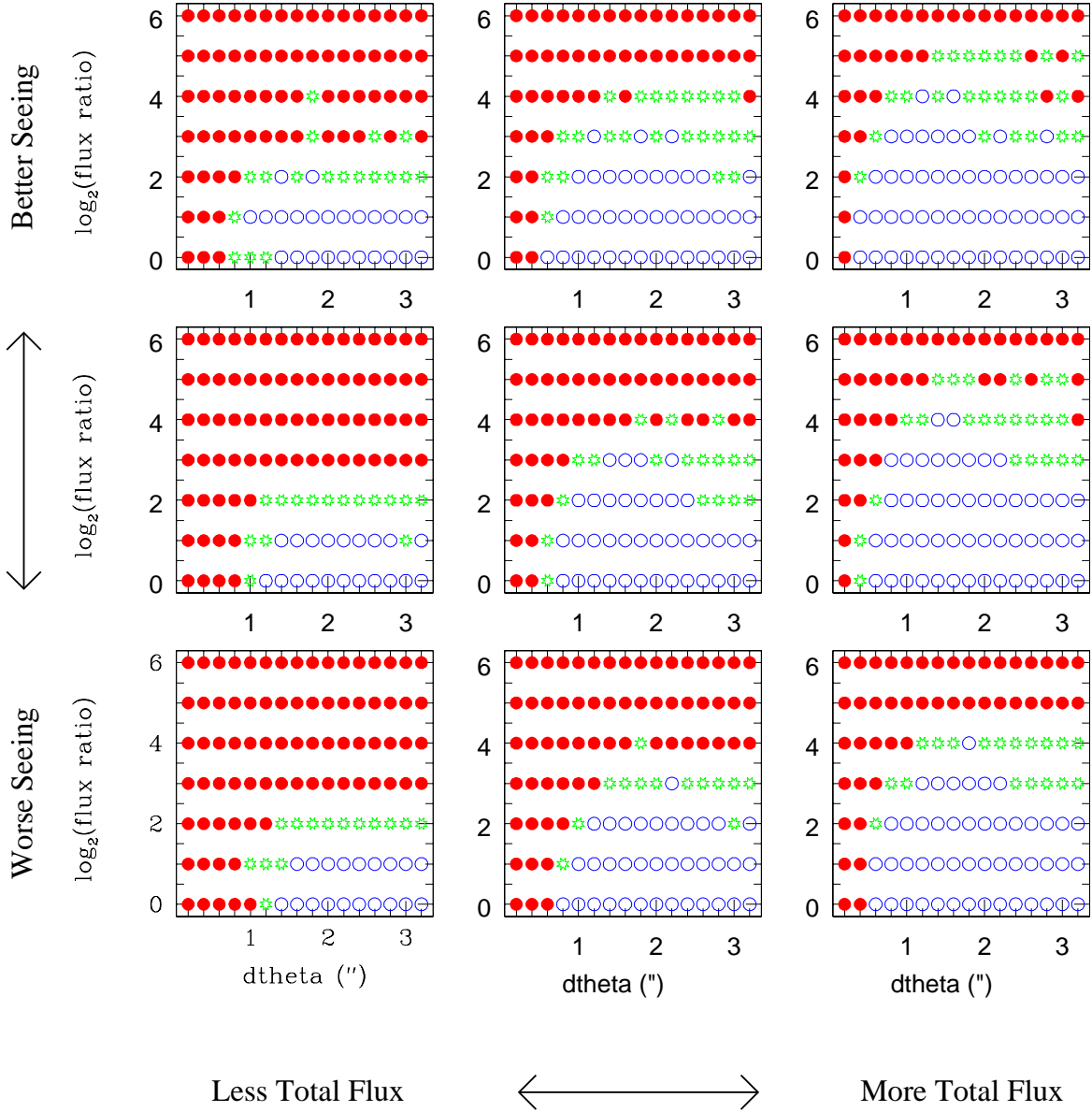,width=\textwidth}
	\caption{Effect of changing the seeing and total flux upon the selection function. The center panel is identical to Figure 2. Going from left to right, the flux decreases by one magnitude (left), is the median (center), increases by one magnitude (right). Going from the bottom to top, the seeing is the 25-th percentile value (1$\farcs$8) (bottom), is the median (1$\farcs$6) (center), is the 75-th percentile value (1$\farcs$4) (top).}

\end{figure}

\begin{figure}

	\centering\epsfig{figure=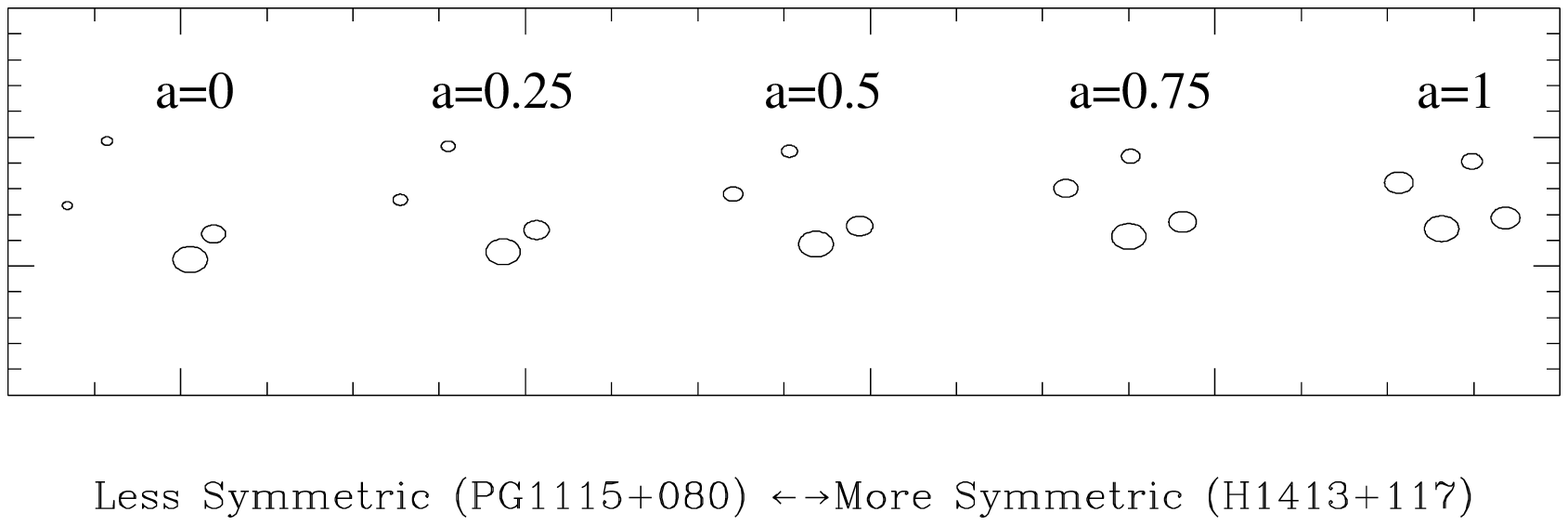,width=\textwidth}
	\caption{Cartoon illustrating a series of five interpolated four-component lens geometries running from less symmetric (PG1115+080,a=0) to more symmetric (H1413+117,a=1). The radii of the circles indicates the relative fluxes of the components.}

\end{figure}

\begin{figure}

	\centering\epsfig{figure=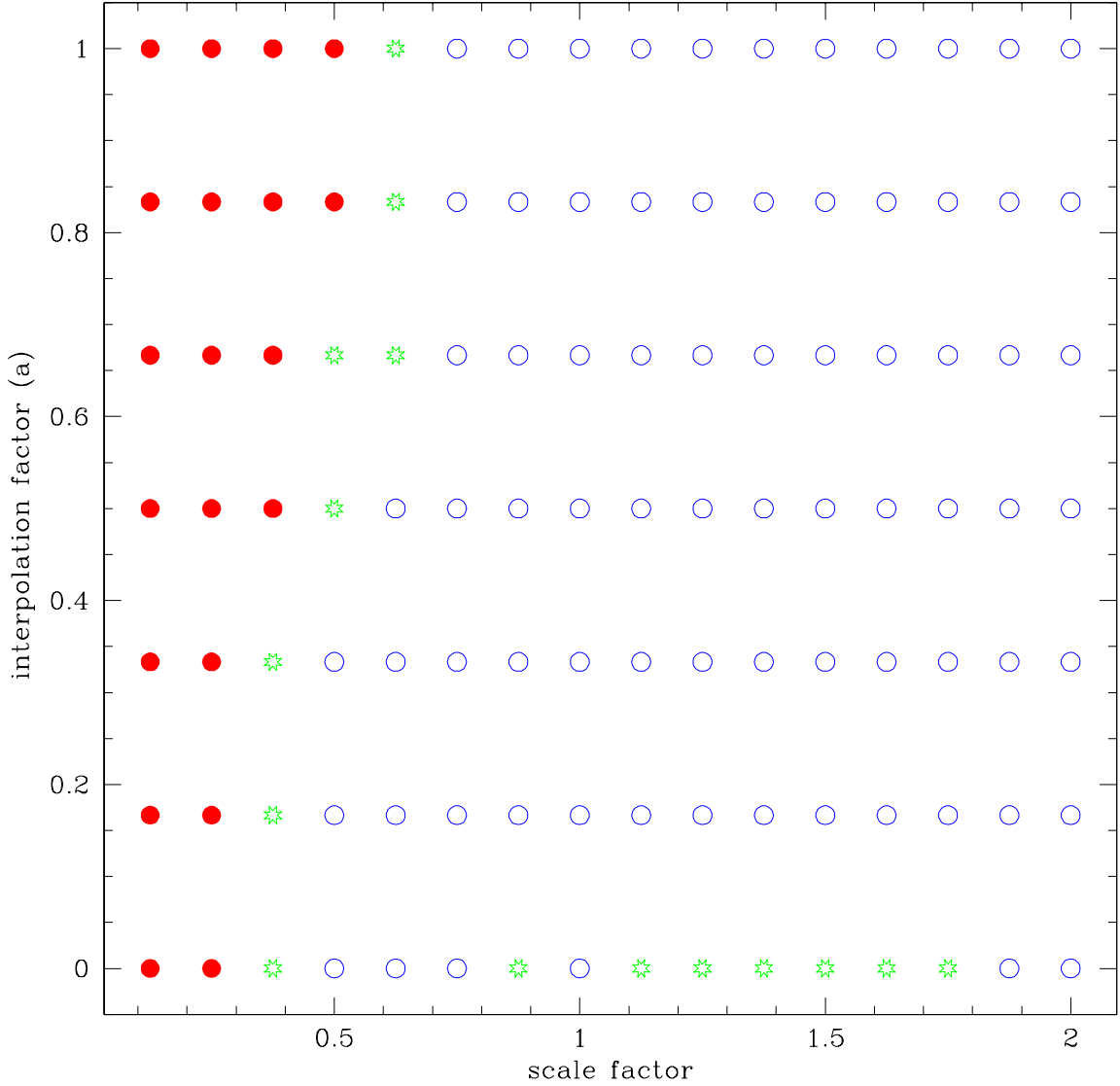,width=\textwidth}
	\caption{Regions of the symmetry - scale  parameter space in which the morphological $\Delta \chi^2 > 0.1 $ cut recovers interpolated quads for the median seeing and flux of SDSS EDR quasars. Along the x-axis the scale of the system is changed between 0.2 and 2.0 of the actual observed size. Along the y-axis the interpolation factor is changed from 0 (asymmetric 1115) to 1 (symmetric 1413). The meaning of the points is the same as in Figure 2.}

\end{figure}

\begin{figure}

	\centering\epsfig{figure=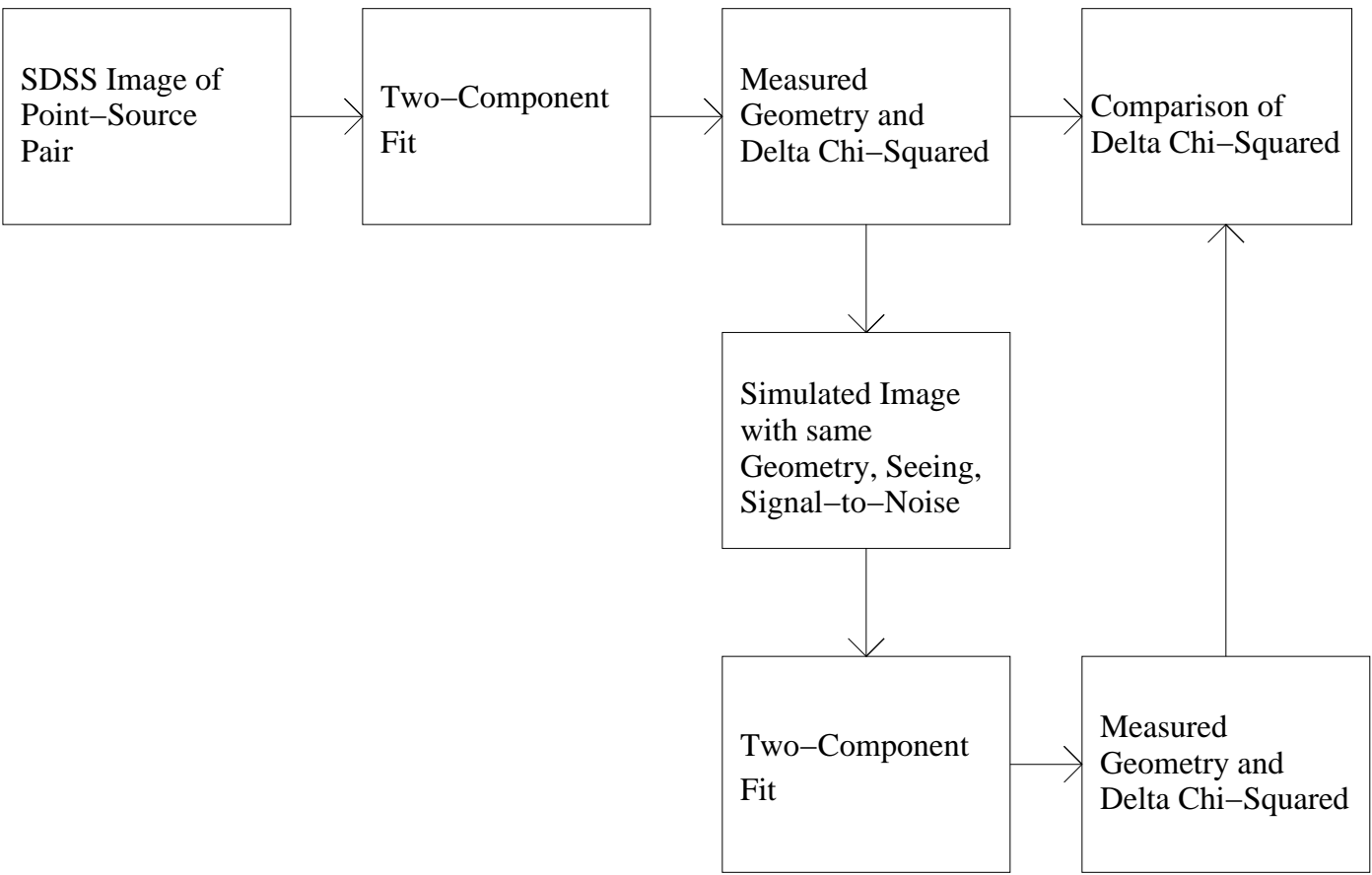,width=\textwidth}
	\caption{Flowchart illustrating our truth-testing procedure for comparing simulated images to real pairs of point-sources from SDSS imaging data}

\end{figure}

\begin{figure}

	\centering\epsfig{figure=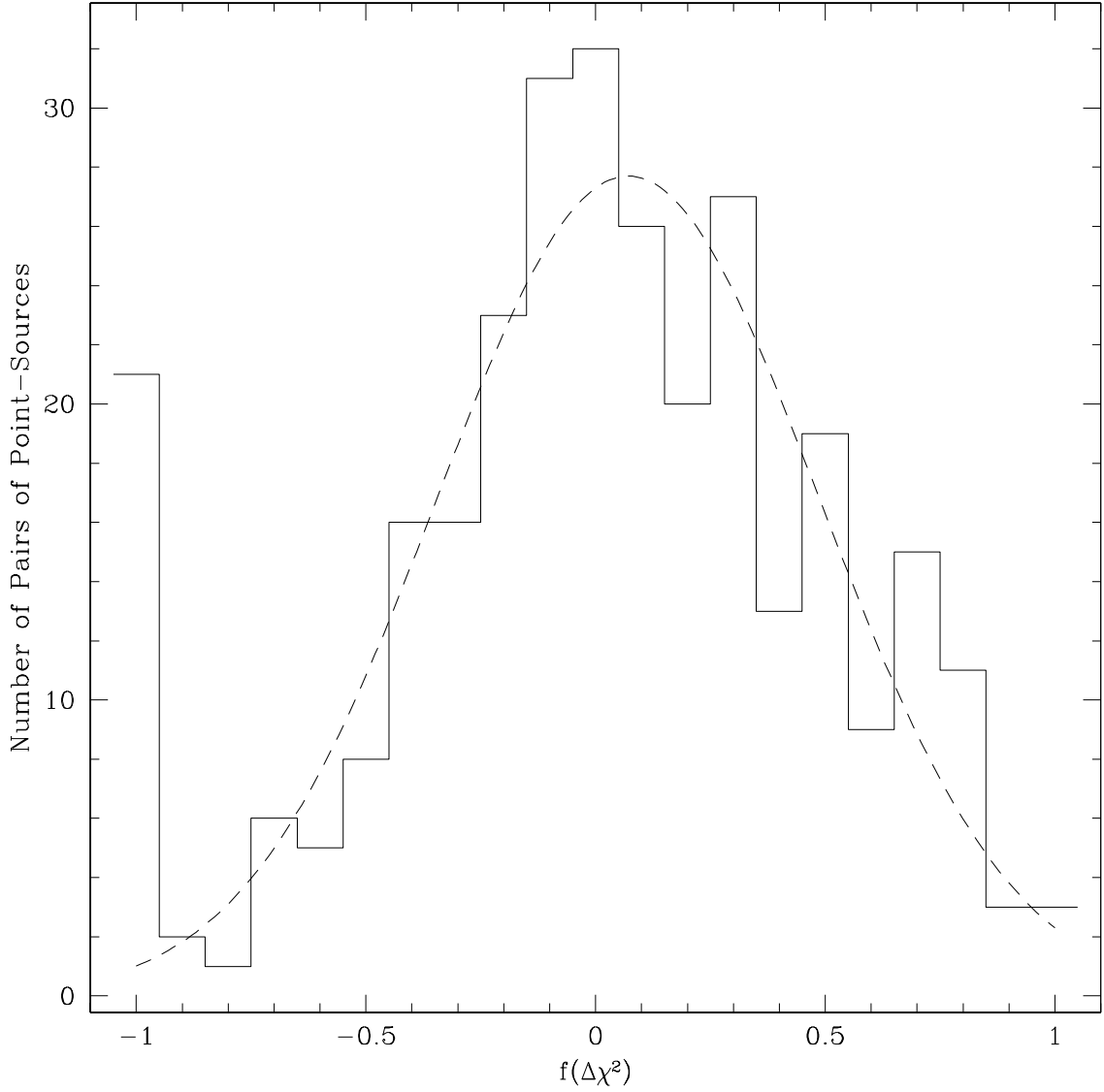,width=\textwidth}
	\caption{Histogram of the distribution of fractional differences in the measured value of $\Delta \chi^2$ between simulated images and real SDSS imaging data. The dashed line is a Gaussian fit to the distribution for values $-1 < f(\Delta \chi^2) < 1$.}

\end{figure}

\begin{figure}

	\centering\epsfig{figure=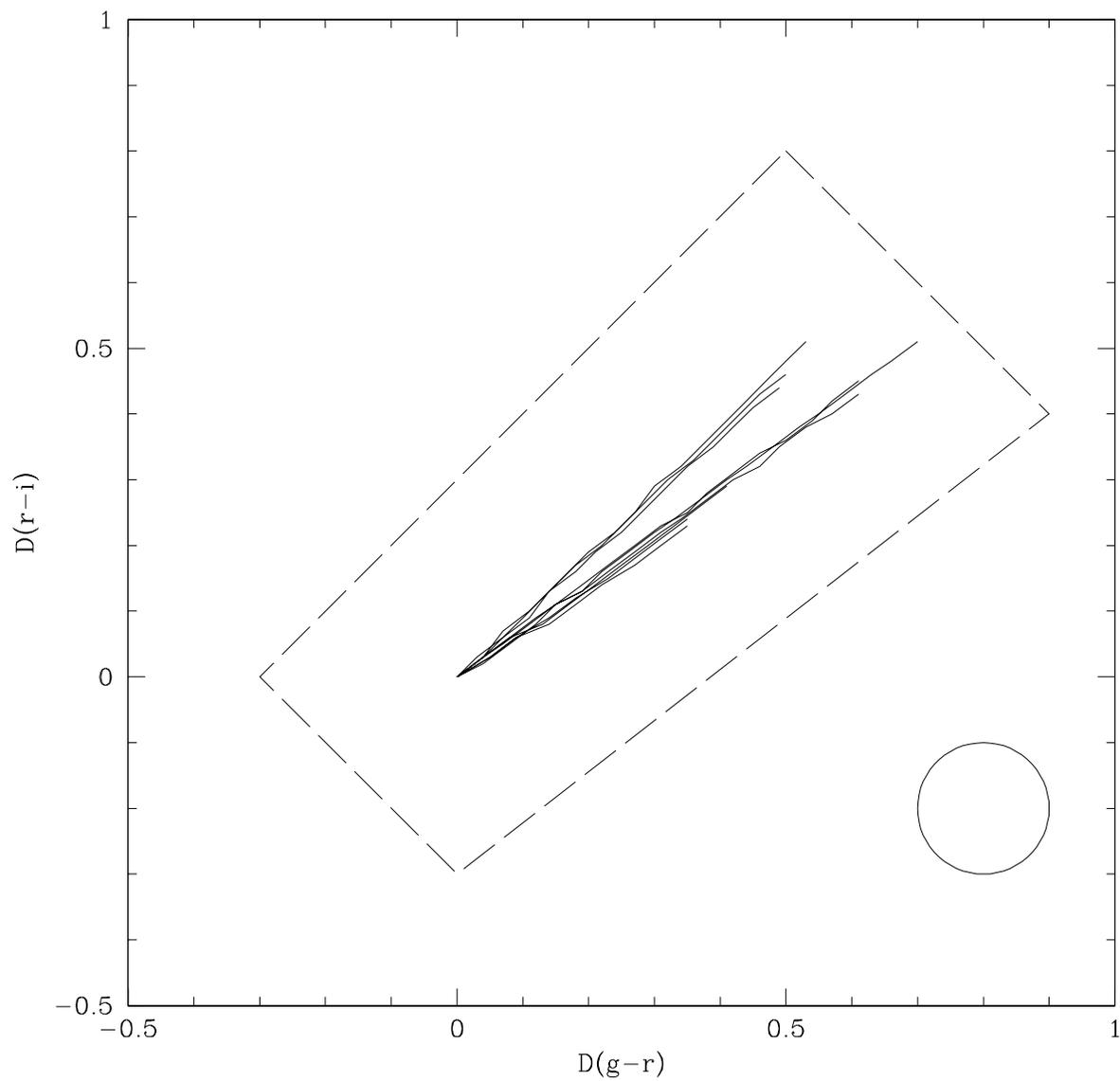,width=\textwidth}
	\caption{Reddening vectors of length $\Delta E(B-V) = 0.5$ for the composite quasar spectrum of Vanden Berk et al. (2001) with quasar redshifts $z_{qso}$ $\in$ \{1 2 3\} and lens galaxy redshifts $z_g$ $\in$ \{0.3 0.5 0.7\} (solid curves). The dashed lines indicate the inclusion region demarked by colour cuts described in eqn. 16 . The circle in the lower right has radius of 0.1, equal to the estimated error in the colour difference measurements, as derived in eqn. 10 .}

\end{figure}

\begin{figure}

	\centering\epsfig{figure=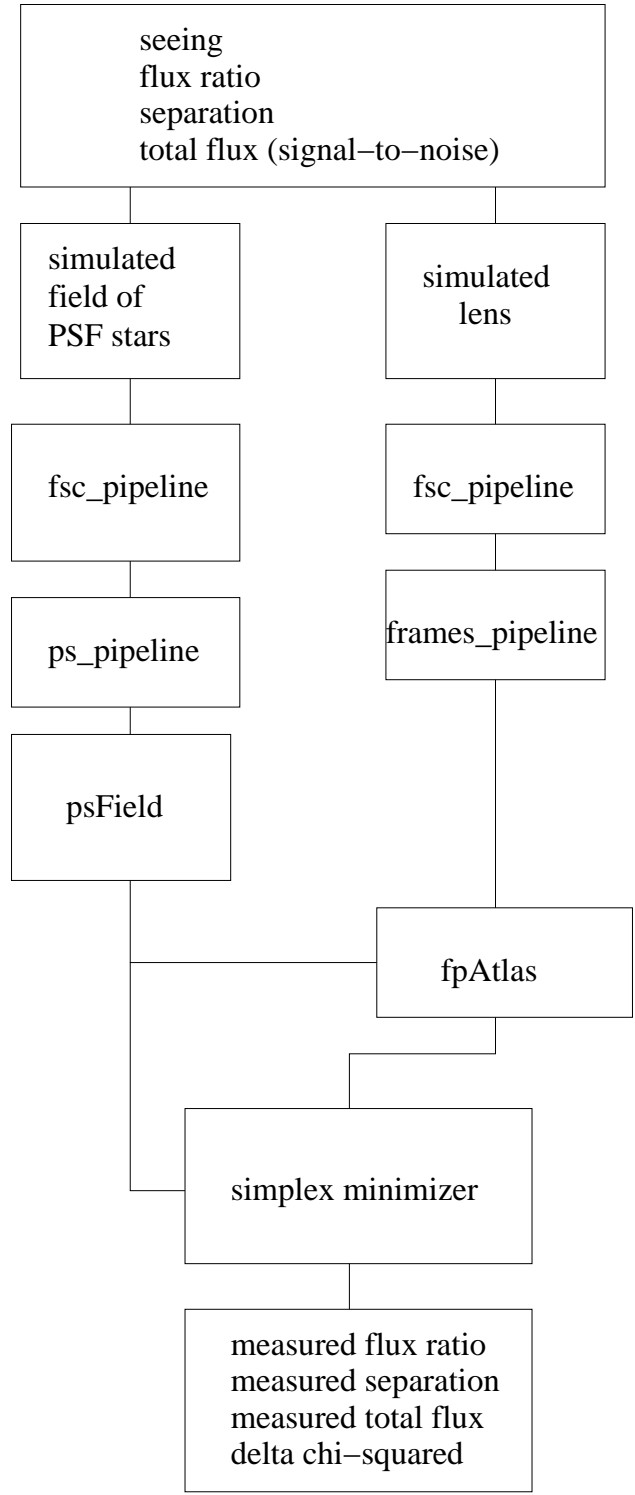,width=\textwidth}
	\caption{Flowchart illustrating the steps in our simulation and psf-fitting procedures. The simulation inputs are in the uppermost box and the psf-fitting outputs in the lower most box.}

\end{figure}

\clearpage

\begin{deluxetable}{cccc}
\tablewidth{0pc}
\tablecaption{Comparison of Model to Data for SDSS1226-0006}
\tablehead{
\colhead{} & \colhead{Band} & \colhead{Separation} & \colhead{Flux Ratio} \\
\colhead{} & \colhead{} & \colhead{($^{\prime\prime}$)} & \colhead{}}
\startdata
\textbf{Observed\tablenotemark{a}} & $g$ & 1.24 $\pm$ 0.02 & 1.8 \\
& $i$ & 1.24 $\pm$ 0.02 & 2.2 \\ 
& & \\
&  $g$ & 1.28 & 1.7 \\
\textbf{Fit to SDSS Data} & $r$ & 1.24 & 1.8 \\
&  $i$ & 1.21 & 2.1 \\ 
\enddata
\tablenotetext{a}{Walter Baade 6.5m with MagIC}
\end{deluxetable}
 
\clearpage

\begin{deluxetable}{ccc}
\tablewidth{0pc}
\tablecaption{Comparison of Simulation to Data for SDSS1226-0006}
\tablehead{
\colhead{} & \colhead{Band} & \colhead{$\Delta \chi^2$}}
\startdata
& $g$ & 7.42 \\
\textbf{Fit to Real SDSS data} & $r$ & 8.34 \\
& $i$ & 5.31 \\
& & \\
& $g$ & 7.64 $\pm$ 0.30 \\
\textbf{Fit to Simulated SDSS data} & $r$ & 10.55 $\pm$ 0.35 \\
& $i$ & 7.01 $\pm$ 0.33  \\
\enddata
\end{deluxetable}

\clearpage

\begin{deluxetable}{cccc}
\tablewidth{0pc}
\tablecaption{Summary of Selection Procedure}
\tablehead{
\colhead{Select On} & \colhead{} & \colhead{Number of} & \colhead{Percentage of}\\
\colhead{} & \colhead{} & \colhead{Remaining Candidates} & \colhead{Candidates Remaining}}
\startdata
Spectrum (QSO $z > 0.6$) & & 5120 & 100 \%\\
$\Delta \chi^2 > 0.1$ & & 408 & 8 \% \\
Separation $<$ 3$^{\prime\prime}$ & & 103 & 2 \% \\
Consistent Geometry $\Delta \mathrm{position} < $ 0.4 & & 41 & 0.8 \% \\
Consistent Colour  & & 13 & 0.3 \% \\ 
\enddata
\end{deluxetable}

\clearpage

\begin{deluxetable}{ccc}
\tablewidth{0pc}
\tablecaption{Selection Results for  Previously-Known Lenses}
\tablehead{
\colhead{Name} & \colhead{Selected by SDSS } & \colhead{Selected as}\\
\colhead{} & \colhead{quasar target selection} & \colhead{Lens Candidate}}
\startdata
APM08279+5255 & N & N \\
SBS0909+532 & Y & Y \\
SDSS0924+0219 & Y & Y \\
SDSS1226-0006 & Y & Y \\
SBS1520+530 & N & N \\
HST15433+5352 & N & N \\
\enddata
\end{deluxetable}

\end{document}